\documentclass[prx,aps,showpacs,twocolumn,superscriptaddress]{revtex4-1}

\usepackage{graphicx,color}

\begin{document}

\title{Multiplicative logarithmic corrections to quantum criticality \\
in three-dimensional dimerized antiferromagnets}

\author{Yan Qi Qin}
\affiliation{Beijing National Laboratory for Condensed Matter Physics and 
\\ Institute of Physics, Chinese Academy of Sciences, Beijing 100190, 
China}
\author{B. Normand}
\affiliation{Department of Physics, Renmin University of China, Beijing 
100872, China}
\author{Anders W. Sandvik}
\affiliation{Beijing National Laboratory for Condensed Matter Physics and 
\\ Institute of Physics, Chinese Academy of Sciences, Beijing 100190, 
China}
\affiliation{Department of Physics, Boston University, 590 Commonwealth 
Avenue, Boston, Massachusetts 02215, USA}
\author{Zi Yang Meng}
\email{zymeng@iphy.ac.cn}
\affiliation{Beijing National Laboratory for Condensed Matter Physics and 
\\ Institute of Physics, Chinese Academy of Sciences, Beijing 100190, China}

\begin{abstract}
We investigate the quantum phase transition in an $S = 1/2$ dimerized 
Heisenberg antiferromagnet in three spatial dimensions. By performing 
large-scale quantum Monte Carlo simulations and detailed finite-size 
scaling analyses, we obtain high-precision results for the quantum 
critical properties at the transition from the magnetically disordered 
dimer-singlet phase to the antiferromagnetically ordered N\'eel phase. 
This transition breaks O($N$) symmetry with $N = 3$ in $D = 3 + 1$ 
dimensions. This is the upper critical dimension, where multiplicative 
logarithmic corrections to the leading mean-field critical properties
are expected; we extract these corrections, establishing their precise 
forms for both the zero-temperature staggered magnetization, $m_s$, and the 
N\'eel temperature, $T_N$. We present a scaling Ansatz for $T_N$, including 
logarithmic corrections, which agrees with our data and indicates exact 
linearity with $m_s$, implying a complete decoupling of quantum and thermal 
fluctuation effects even arbitrarily close to the quantum critical point. 
We also demonstrate the predicted $N$-independent leading and subleading 
logarithmic corrections in the size-dependence of the staggered magnetic 
susceptibility. These logarithmic scaling forms have not previously been 
identified or verified by unbiased numerical methods and we discuss their 
relevance to experimental studies of dimerized quantum antiferromagnets 
such as TlCuCl$_3$.
\end{abstract}

\pacs{75.10.Jm, 75.40.Cx, 75.40.Mg}

\date{\today} 

\maketitle

\section{Introduction}

Antiferromagnetic insulators exhibit a multitude of fundamental phenomena 
in the neighborhood of the phase transitions separating their magnetically 
ordered ground states from different types of quantum paramagnetic phase. 
These quantum phase transitions (QPTs) occur at temperature $T = 0$ as a 
consequence of non-thermal parameters (examples include magnetic fields, 
applied pressure, and dopant concentration) that act to change the effect 
of quantum mechanical fluctuations \cite{Sachdev2008,Kaul2013}. At finite 
temperatures, a further dimension is opened in the presence of both quantum 
and classical (thermal) fluctuations, and the rich physics arising from their 
interplay includes all the properties of the quantum critical (QC) regime 
\cite{rs}. 

Experimentally, the material in which the most detailed study of intertwined 
classical and quantum critical behavior has been performed is TlCuCl$_3$. 
This compound is composed of antiferromagnetically coupled pairs of 
Cu$^{2+}$ ions ($S = 1/2$), which tend to form dimer singlets and have 
antiferromagnetic interdimer couplings in all three spatial dimensions 
($d = 3$) \cite{Matsumoto2004}. At ambient pressure and zero field, 
TlCuCl$_3$ is a nonmagnetic insulator with a gap of 0.63 meV to triplet 
spin excitations. As a consequence of this small gap, an applied magnetic 
field of 5.4 T is sufficient to drive the system to an ordered 
antiferromagnetic state, through a QPT in the Bose-Einstein universality 
class \cite{rgrt}. A relatively small applied hydrostatic pressure, 
$p_c = 1.07$ kbar, is also sufficient to create an antiferromagnetically 
ordered state \cite{rrs2004}, through a QPT in the three-dimensional (3D) 
O(3) universality class due to spontaneous breaking of the SU(2) spin 
symmetry (which is further reduced in TlCuCl$_3$ by a weak unixial 
anisotropy, making the universality class 3D XY). 

The elementary excitations on the ordered side of the zero-field quantum
critical point (QCP) are gapless spin waves, the Goldstone modes associated 
with spontaneous breaking of spin-rotational symmetry. On the disordered side, 
quantum fluctuations, towards spin-singlet formation on the dimers, suppress 
the long-range antiferromagnetic order, restoring the symmetry and ensuring 
that all excitations are gapped. This evolution of the excitation spectrum 
in TlCuCl$_3$ has been measured in Ref.~\cite{Ruegg2008}. At finite 
temperatures on the ordered side, a classical phase transition occurs at the 
N\'eel temperature, $T_{N}$, where the long-ranged magnetic order is ``melted'' 
not by quantum but by thermal fluctuations. At finite temperatures around the 
QCP, the combination of quantum and thermal fluctuations creates the QC 
regime, where the only characteristic energy scale of the system is the 
temperature itself and many universal properties emerge \cite{rs}. The phase 
diagram of TlCuCl$_3$ under pressure and the restoration of classical critical
scaling around $T_{N}$ were the subject of a recent investigation 
\cite{Merchant2014}. 

\begin{figure}[t]
\includegraphics[width=8cm]{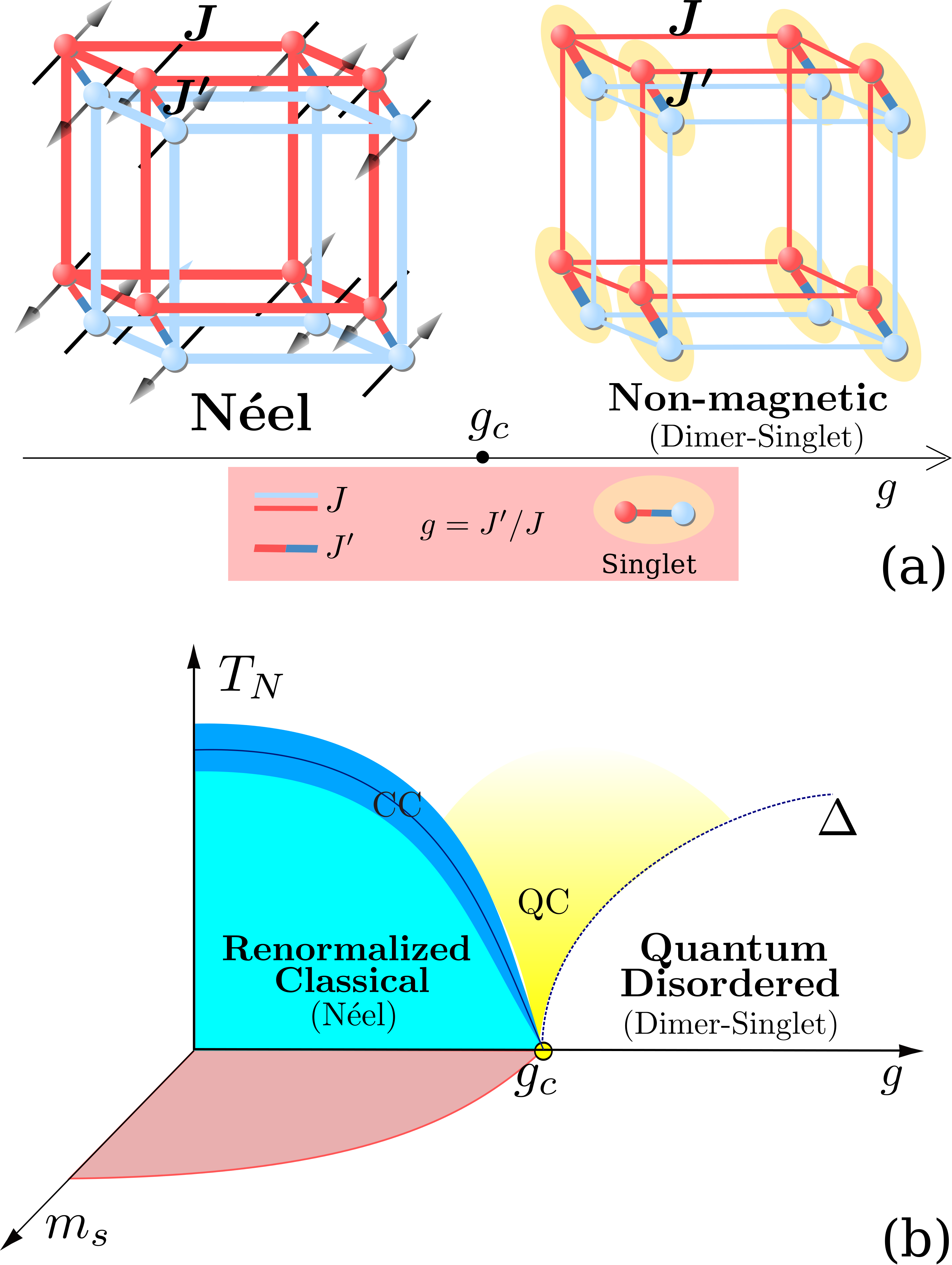}
\caption{(Color online) (a) Dimerized lattice of $S = 1/2$ spins in the 3D 
double cubic geometry. Sites of the red and blue cubic lattices are connected 
pairwise by dimer bonds; $J'$ and $J$ are antiferromagnetic Heisenberg 
interactions, respectively on and between the dimer units, and their ratio, 
$g = J'/J$, controls the QPT from a N\'eel ordered phase (left) to a quantum 
disordered dimer-singlet phase (right), with the QCP occurring at the critical 
ratio $g_{c}$. (b) Schematic quantum critical phase diagram for the Heisenberg 
model on the double cubic lattice. On the ordered ``renormalized classical'' 
side, $g < g_c$, the N\'eel order is progressively weakened by increasing 
quantum fluctuations as $g$ approaches $g_{c}$, causing both the ordering 
(N\'eel) temperature, $T_{N} (g)$, and the order parameter (the staggered 
magnetization), $m_{s} (g)$, to go continuously to zero. On the ``quantum 
disordered'' side, intradimer correlations dominate and the characteristic 
energy scale, $\Delta$, is the gap to triplet excitations. Above $g_c$ at
$T > 0$ is the universal ``quantum critical'' (QC) region, whose behavior 
is governed by the $3 + 1$-dimensional O(3) universality class. Around $T_N 
(g)$, one expects a region of classical critical (CC) behavior where thermal 
fluctuations are dominant.}
\label{fig:lattice}
\end{figure}

QPTs in dimerized quantum spin models have been studied numerically by a 
number of authors, primarily by quantum Monte Carlo (QMC) simulations. Early 
investigations of the bilayer square-lattice antiferromagnet \cite{Sandvik1994}
and other two-dimensional geometries \cite{Troyer1996,Matsumoto2001} have 
been followed more recently by high-precision studies of a range of critical 
properties \cite{Wang2006,Meng2008,Wenzel2009,Sandvik2010,Sen2015}. In three 
dimensions, the focus of investigations has been on the field-induced 
transition \cite{Nohadani2005}, on the effects of dimensionality 
\cite{Troyer1997,Yao2007}, and on physical observables at the 
coupling-induced QPT \cite{Jin2012,Oitmaa2012,Kao2013}. 

A minimal model of a dimerized quantum antiferromagnet has only 
two coupling constants, $J'$ on and $J$ between the dimer units, and therefore 
only one control parameter, $g = J'/J$. The geometry considered in the present 
study is the double cubic lattice shown in Fig.~\ref{fig:lattice}(a). In this 
system at large $g$, intradimer singlet correlations dominate the physics and 
the ground state is magnetically disordered, while at small $g$ the interdimer 
correlations establish long-ranged magnetic order. The order parameter of the 
N\'eel phase is the staggered magnetization, $m_s(g)$, and along with the 
ordering temperature, $T_N (g)$, it can be driven continuously to zero by 
increasing $g$, as illustrated in Fig.~\ref{fig:lattice}(b). By standard 
arguments of dimensionality and symmetry, the dynamical exponent of this 
system is $z = 1$ \cite{rs,Chakravarty88} and the QCP belongs to the $D = 
3 + 1$ O(3) universality class, which is at the upper critical dimension, 
$D_{c} = 4$, of all O($N$) models \cite{Zinn-Justin2002,rs}. At $D = D_c$, 
mean-field critical scaling behavior alone is not sufficient to capture 
the physics of fluctuations around the QCP and multiplicative logarithmic 
corrections to the physical quantities (thermodynamic functions) are expected.

The theoretical importance of multiplicative logarithmic corrections to 
mean-field scaling behavior lies not only in the statistical physics of 
condensed matter systems but also in high-energy physics \cite{Zinn-Justin2002,
rkbc}. The general problem of a quantum field theory with an $N$-component 
field is encapsulated by an ``O($N$) $\phi^{4}$ theory,'' a Lagrangian 
containing a dynamic (quadratic gradient) term and a potential term with 
quadratic ($\phi^2$) and quartic ($\phi^4$) contributions. On changing the 
sign of the quadratic term, the system is driven through a QPT separating a
phase with $\langle \phi \rangle = 0$ from one with $\langle \phi \rangle \ne 
0$ (a ``Mexican hat'' potential). As noted above, the low-energy properties 
of the 3D dimerized antiferromagnet of SU($2$) quantum spins with Heisenberg 
interactions correspond to a field theory with $N = 3$ and $D = 3 + 1$ 
(including the time dimension); $N = 1$ and $2$ correspond respectively 
to Ising and XY spin interactions. 

Beyond the upper critical dimension $(D > D_{c})$, the scaling behavior of 
the O($N$) $\phi^4$ theory is straightforward, with the critical exponents 
being exactly those given by mean-field theory \cite{Aizenman1981,rbg,rkbc}, 
namely $\alpha = 0$, $\beta = {1}/{2}$, $\gamma = 1$, $\delta = 3$, and 
$\nu = {1}/{2}$. As we discuss below, this may be taken as an expression 
of the independence of quantum and thermal fluctuations when the phase space 
is sufficiently large. For $D < D_c$, the situation is complex and these 
exponents take anomalous values. However, exactly at the upper critical 
dimension, $D = D_{c} = 4$, the leading scaling behavior coincides with 
that of the mean-field theory, but modified by multiplicative logarithmic 
corrections \cite{rkl,rkle,Zinn-Justin2002}. While the leading exponents 
are $N$-independent, a measure of $N$-dependence resides in the logarithmic 
corrections and these must be taken into account to establish the universality 
class of the transition \cite{Janssen2004}. Because the established results 
for the form of these corrections are based on perturbative techniques applied 
to low-energy theories, it is desirable to verify them using unbiased numerical
methods applied directly to the lattice Hamiltonians, and this is what we 
achieve here. 

Despite the insight into general QPT phenomena obtained from simulations 
using this type of minimal model for 3D dimerized systems \cite{Nohadani2005,
Troyer1996,Yao2007,Jin2012,Oitmaa2012}, the question of logarthmic corrections 
has to date been addressed only briefly and inconclusively \cite{tbk07,Kao2013}.
Experimentally, the feasibility of observing logarithmic corrections in systems
such as TlCuCl$_3$ remains a challenging open issue \cite{Merchant2014}. In 
this paper we provide a systematic numerical study. We employ large-scale QMC 
simulations to investigate the critical behavior of the order parameter and 
N\'eel temperature on the double cubic lattice [Fig.~\ref{fig:lattice}(a)] 
for small values of $|g - g_c|$ unattainable in all previous studies. 
State-of-the-art QMC techniques and finite-size-scaling analysis using 
very large systems (sizes exceeding $10^5$ spins) allow us to detect and 
characterize the multiplicative logarithmic corrections in the universal 
scaling relations for the QC regime at the upper critical dimension, here 
for the $D = 3 + 1$ O(3) QCP. In fact our results constitute hitherto 
unavailable exact numerical verification of the logarithmic forms predicted 
both by perturbative renormalization-group calculations \cite{rw,rwr,rblzj} 
based on the $N$-component $\phi^{4}$ theory at $D_c$ and by additional 
considerations exploiting the zeros of the partition function 
\cite{rkl,rkle,Kenna2004,rkbc}. To the best of our knowledge, no systematic
numerical calculations have been performed beyond $N = 1$ \cite{rkl,rfkp}. 

As will become clear, our numerical results demonstrate to high precision 
the validity of the detailed theoretical predictions for the expected 
universality class. Both size-dependent scaling and the order parameter in 
the thermodynamic limit show evident deviations from pure mean-field behavior, 
which are accounted for by logarithmic corrections whose exponents are in very 
close ageement with the predicted values where available. In the case of the 
N\'eel temperature, we are not aware of any previous scaling predictions 
including logarithms. Here we test an Ansatz based on the known scaling 
behavior for the relevant energy scales \cite{rs,Fisher1989} and the 
logarithmic corrections in corresponding classical systems \cite{rkbc}.
In addition to probing the asymptotic behavior, our results also provide 
direct insight into the range of validity of logarithmically modified 
critical scaling forms as one moves away from the QCP, which will be 
essential in evaluating the experimental relevance of logarithmic corrections.

The paper is organized as follows. In Sec.~\ref{sec:methods} we introduce 
the model and the numerical method, describing the measurement of physical 
observables in our QMC simulations. In Sec.~\ref{sec:qcp} we begin the 
presentation  of our numerical results with the precise determination of 
$g_c$, the position of the QCP, using finite-size-scaling techniques. 
Section \ref{sec:chilogs} discusses the observation of clear logarithmic 
corrections in the staggered magnetic susceptibility, ${\chi}(\mathbf{Q}
_{\rm AF},L)$, at the QCP as a function of the system size, $L$. We present 
our results for the sublattice magnetization, $m_s$, at $T = 0$ in 
Sec.~\ref{sec:ms}, discussing in detail its extrapolation to the 
thermodynamic limit, where we investigate the presence of logarithmic 
corrections to the leading mean-field behavior. In Sec.~\ref{sec:tn} we 
present a scaling Ansatz for the N\'eel temperature, apply finite-size 
scaling to extract it as a function of $g$, and again investigate corrections 
to mean-field behavior. We compute the characteristic velocity $c$ of spin 
excitations, demonstrate the precise linearity of $T_N$ and $m_s$, and 
discuss the physical interpretation of this behavior. We summarize our 
results in Sec.~\ref{sec:summary} and comment further on their theoretical 
and experimental consequences.

\section{Model and methods}
\label{sec:methods}

As a representative 3D dimerized lattice with an unfrustrated geometry, we 
choose to study the double cubic model shown in Fig.~\ref{fig:lattice}(a). 
This system consists of two interpenetrating cubic lattices with the same 
antiferromagnetic interaction strength, $J$, connected pairwise by another 
antiferromagnetic interaction, $J'$. The QPT occurs when the coupling ratio 
$g = J'/J$ is increased, changing the ground state from a N\'eel-ordered phase 
of finite staggered magnetization to a dimer-singlet (``quantum disordered'') 
phase, as illustrated in Fig.~\ref{fig:lattice}(b). An advantage of this 
geometry over cases where the dimerization is imposed within a single lattice, 
such as the simple cubic lattice \cite{Jin2012}, is that all symmetries of 
the cubic lattices are retained, facilitating the consideration of quantities 
such as the spin stiffness or the velocity of spin excitations.

The Hamiltonian is given by
\begin{equation}
H = \sum_{\langle i,j \rangle} J_{ij} {\vec S}_i \cdot {\vec S}_j,
\label{eq:hamiltonian}
\end{equation}
where ${\vec S}_i$ is an $S = 1/2$ spin operator residing on a double cubic 
lattice of $N = 2L^{3}$ sites with periodic boundary conditions. The sum is 
taken only over nearest-neighbor sites, where every site has six neighbors 
on the same cubic lattice with coupling strength $J_{ij} = J$ and one neighbor 
on the opposite cubic lattice with $J_{ij} = J'$ [Fig.~\ref{fig:lattice}(a)]. 
We set $J = 1$ as the unit of energy and use $g = J'/J$ as the control 
parameter. 

To study this system, we use the stochastic series expansion (SSE) QMC 
technique \cite{Sandvik1991,Sandvik1999,Evertz2003,Sandvik2010} to obtain 
unbiased results, i.e.~numerically exact within well-characterized statistical 
errors, for physical quantities in systems of finite side-length $L$. Here we 
present results up to $L = 48$ at temperatures $T = \beta^{-1}$ with $\beta$ 
up to $2L$. We then perform detailed analyses by finite-size scaling 
\cite{Fisher1972} to extract information in the thermodynamic limit both in 
the ordered state and at the QCP, as detailed in the separate sections to 
follow. Here we define the physical quantities of interest and discuss some 
technical aspects of their calculation within the SSE QMC method.

Because spin-rotation symmetry is not broken in simulations of finite-size 
systems, one may measure the squared order parameter and take its square 
root as a post-simulation step. The staggered magnetization is given by 
\begin{equation}
m_{s} = \sqrt{\frac{1}{N} S(\mathbf{Q}_{\rm AF})},
\label{esm}
\end{equation}
where 
\begin{equation}
S(\mathbf{q}) = \frac{1}{N} \sum_{i,j}^N e^{-i\mathbf{q} \cdot (\mathbf{r}_{i} - 
\mathbf{r}_{j})} \langle \vec{S}_i \cdot \vec{S}_j \rangle
\label{emsf}
\end{equation}
is the magnetic structure factor, with $\mathbf{r}_{i}$ denoting the real-space 
position of the spin $\vec{S}_i$ on lattice site $i$, and $\mathbf{Q}_{\rm AF}
 = (\pi,\pi,\pi,\pi)$ is the wave vector of antiferromagnetic order, with the 
fourth $\pi$ denoting the phase factor between the two cubic lattices. We 
consider only the $z$-component of the magnetization and average it over the 
time dimension of the QMC configurations, computing the expectation value of 
the squared order parameter in the form
\begin{equation}
m^2_{sz} = \frac{1}{\beta} \int^\beta_0 d\tau m^2_{sz}(\tau), 
\label{m2sztau}
\end{equation}
where
\begin{equation}
m_{sz} (\tau) = \frac{1}{N} \sum_{i=1}^N e^{-i\mathbf{Q}_{\rm AF} \cdot 
\mathbf{r}_{i}} S^z_i (\tau),
\end{equation}
with 
\begin{equation}
S^z_i (\tau) = e^{\tau H} S^z_i e^{-\tau H}
\end{equation}
the time-evolved spin operator at imaginary time $\tau$. In an SSE simulation, 
the integral in Eq.~(\ref{m2sztau}) is transformed into a discrete sum with no 
approximations and the relation compensating for the rotational averaging 
of the single measured component of the order parameter,
\begin{equation}
m_s = \sqrt{3\langle m^2_{sz}\rangle},
\label{ms3mz}
\end{equation}
is applied post-simulation. The magnetic susceptibility is defined as
\begin{equation}
{\chi} (\mathbf{q}) = \frac{1}{N} \sum_{ij} \int^\beta_0 d\tau\langle 
S^z_i(\tau) S^z_j(0) \rangle e^{-i\mathbf{q} \cdot (\mathbf{r}_i - \mathbf{r}_j)}.
\label{ems}
\end{equation}
In the SSE approach, the squared order parameter [Eq.~(\ref{m2sztau})] 
is readily evaluated at any $\tau$ because the QMC configurations are 
constructed in the $S^z$ basis \cite{Sandvik1997,Sandvik2010}. The dynamical 
spin-spin correlation function contained in Eq.~(\ref{ems}) can also be 
obtained easily by applying an operator string connecting the $S^z$ states 
at different imaginary times, with the integral computed analytically to 
give a direct, formally exact QMC estimator not requiring post-simulation 
integration \cite{Sandvik1999,Sandvik2010}.

The Binder ratio \cite{Binder1981} is the ratio of the fourth moment of a 
quantity to the square of its second moment. For our purposes, the relevant 
quantity is 
\begin{equation}
R_2 = \frac{\langle m_{sz}^4 \rangle}{\langle m_{sz}^2 \rangle^2},
\label{er2}
\end{equation}
which is dimensionless and satisfies the crucial property of being 
size-independent at the QCP in the limit of large system sizes. The spin 
stiffness, or helicity modulus, of the system is defined as
\begin{equation}
\rho^\alpha_s = \frac{1}{N} \frac{\partial^2 F (\phi_\alpha)}{\partial^2 
\phi_\alpha}\Big |_{\phi_\alpha \to 0}, \;\;\;\;\;\; \alpha = x,y,z,
\label{RhosFphi}
\end{equation}
where $F$ is the free energy and $\phi_\alpha$ is the angle of a twist imposed 
between all spins in planes perpendicular to the $\alpha$ axis. In an SSE 
simulation, the most efficient way to extract the spin stiffness is to take 
the derivative in Eq.~(\ref{RhosFphi}) directly in the QMC expression for 
$F(\phi_\alpha)$ at $\phi_\alpha = 0$, giving
\begin{equation}
\rho^\alpha_s = \frac{3 \langle w^{2}_{\alpha} \rangle}{4 \beta},
\label{ess}
\end{equation}
where 
\begin{equation}
w_{\alpha} = \frac{1}{L} (N^{+}_{\alpha} - N^{-}_{\alpha})
\label{ewn}
\end{equation} 
is the winding number \cite{Pollock1987,Sandvik1997} of the spin in spatial 
direction $\alpha$ and $N^{+}_{\alpha}$ and $N^{-}_{\alpha}$ are the numbers of 
occurrences of the operators $S^{+}_{i} S^{-}_{j}$ and $S^{-}_{i} S^{+}_{j}$ on 
bond $\langle i,j \rangle$ in the $\alpha$-direction within imaginary time 
$[0,\beta]$. As noted above, $\rho^\alpha_s$ is the same in all three directions
due to the symmetry of the double cubic lattice, and the average may be taken 
over all of these. The spin stiffness follows the scaling law $\rho_s \propto 
L^{2-d-z}$ in $d$ spatial dimensions \cite{Sandvik2010} and, because the 
dynamic exponent is $z = 1$ here, the quantity $\rho_s L^{d-1}$, or 
equivalently $\rho_s L^{D-2}$, is also size-independent at the QCP, up to 
a logarithmic correction at the upper critical dimension. 

Finally, the spin-wave velocity, $c$, can be obtained reliably by 
monitoring the fluctuations of the spatial and temporal winding numbers 
\cite{Kaul2008,Jiang2011,Kao2013,Sen2015}. For a fixed system size, $L$, 
the inverse temperature $\beta$ is adjusted to the value $\beta^*$ where 
the system has equal winding-number fluctuations in the spatial and temporal 
directions,
\begin{equation}
\langle w^{2}_{\alpha}(\beta^{*}) \rangle = \langle w^{2}_{\tau}(\beta^{*}) 
\rangle, \;\;\;\;\;\; \alpha = x,y,z.
\label{winding}
\end{equation}
The temporal winding number is the net magnetization (number of up minus 
number of down spins) of the system, $w_\tau = M_z = \sum_i S_i^z$, which 
is easily obtained in the $S^z$ basis \cite{Sen2015}. When the condition 
(\ref{winding}) is met, the spin-wave velocity is given by the ratio
\begin{equation}
c = \frac{L}{\beta^{*}(L)}.
\label{spinwavevelocity}
\end{equation}
The isotropy of the lattice is an advantage also in this case. For each 
value of $g$, an extrapolation $L \to \infty$ is performed to obtain $c$ 
in the thermodynamic limit. For further details of these procedures, we 
refer the reader to the recent extensive tests of this method conducted in 
Ref.~\cite{Sen2015}.  

\section{Determination of the QCP}
\label{sec:qcp}

The key to an accurate characterization of logarithmic corrections is a 
high-precision determination of the location $g_c$ of the QCP. For this we 
employ the Binder ratio, $R_{2}$, and the appropriately scaled spin stiffness, 
$\rho_s L^{D-2}$, which both have scaling dimension zero and therefore should 
approach constant values at $g_c$ when $L \to \infty$, up to possible 
logarithmic corrections. We stress that the scaling forms for the approach 
of both quantities to the critical point are valid for a four-dimensional 
(4D) theory, with the temperature (imaginary time) providing the fourth 
dimension, and are applicable on a ``critical contour'' where the inverse 
temperature $T^{-1} = kL$ is taken to infinity symmetrically with the spatial 
dimension of the system. This form is appropriate for a system with dynamic 
exponent $z = 1$ (in general $1/T \sim L^z)$. All values of $k$ yield the 
same results in the limit $L \to \infty$, and in principle the contour is 
optimal when $k = 1/c$; the spin-wave velocity, $c$, is a number of order 
unity discussed in detail in Sec.~\ref{sec:tn}, and here we use $k = 1$. 

Away from $g_c$, $R_{2}$ and $\rho_s L^{2}$ approach different constant values 
with increasing system size. In the N\'eel state, $R_2$ approaches $9/5$ due to 
diminishing fluctuations in the magnitude of the rotationally-invariant order 
parameter, of which we measure only the $z$-component in Eq.~(\ref{er2}). In 
the quantum disordered phase, $R_2$ approaches a higher value dictated by 
Gaussian fluctuations, which from the symmetries of the double cubic model is 
$3$. The spin stiffness falls from non-zero values in the N\'eel phase to zero 
in the disordered phase. When calculated as functions of $g$, the curves 
$R_{2}(g,L)$ and $\rho_s(g,L)L^{2}$ obtained for different system sizes should 
cross at the QCP, up to corrections that are well understood from the theory of 
finite-size scaling. We analyze these corrections to obtain an unbiased value 
of the critical coupling, $g_c$, in the thermodynamic limit \cite{Sandvik2010}. 

\begin{figure}[t]
\centering
\includegraphics[width=8.4cm]{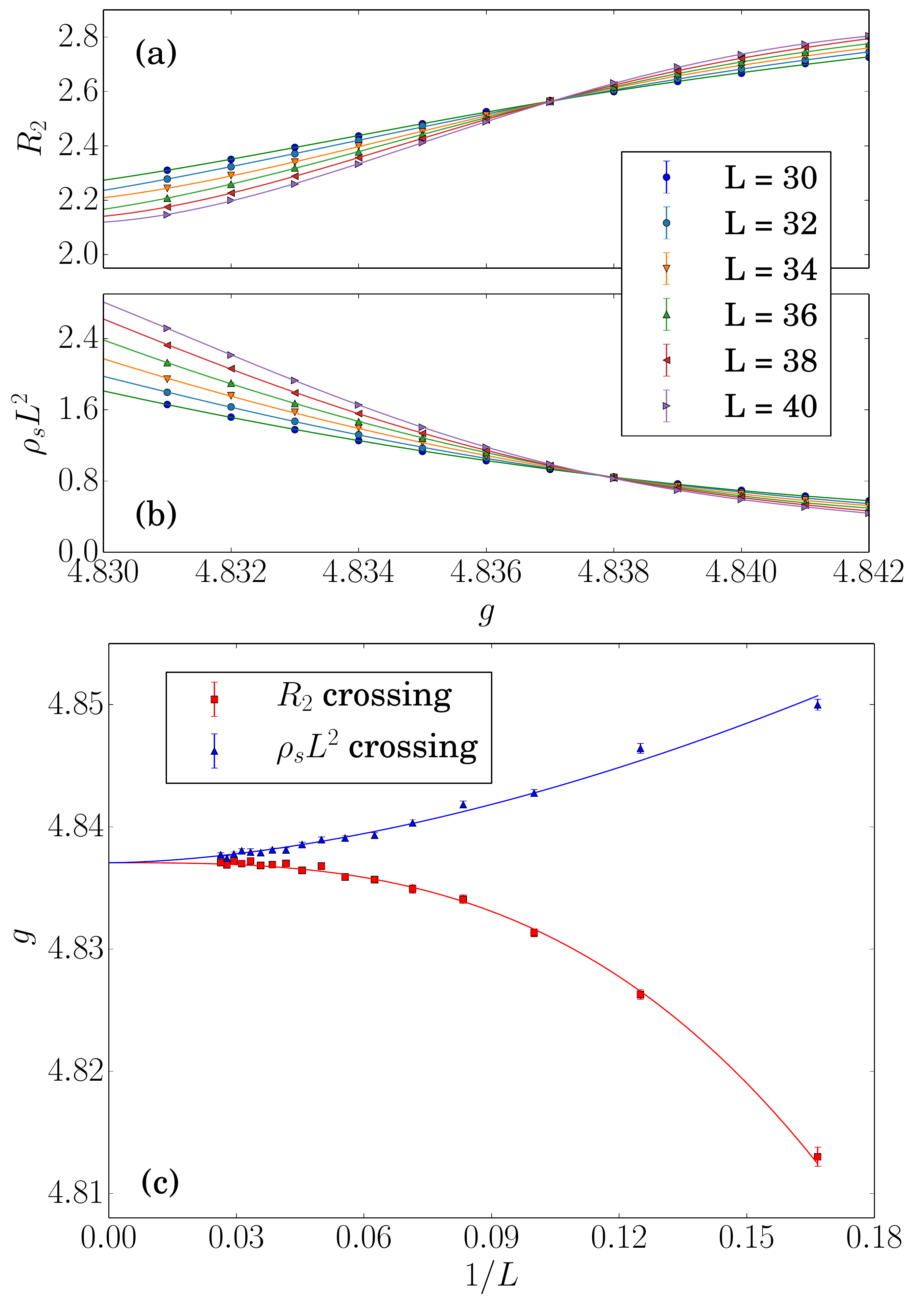}
\caption{(Color online) (a) Binder ratio, $R_{2}$, and (b) scaled spin 
stiffness, $\rho_{s}L^2$, as functions of the coupling ratio $g$ for system 
sizes $L = 30$, 32, $\dots$, 40. Crossings of the curves for pairs of system 
sizes $L$ and $L+2$ define finite-size estimates $g_{c}^R (L)$ and $g_{c}^{\rho} 
(L)$ of the critical point, which are are fitted to the form of 
Eq.~(\ref{gcpowerform}) in panel (c). Enforcing a common value of $g_c$ in 
both fits gives the $L \to \infty$ critical point as $g_{c} = 4.83704(6)$ 
and the irrelevant exponent as $\omega = - 0.31(5)$ for $R_2$ and $\omega
 = 0.82(5)$ for $\rho_{s}L^2$.}
\label{fig:fssofg}
\end{figure}

Figure~\ref{fig:fssofg}(a) shows $R_2$ as a function of $g$ in the 
neighborhood of the critical coupling ratio for various system sizes. We 
have performed simulations for all even-length sizes $L = 6$, 8, 10, \dots, 
40, but here we present only the $L = 30$, 32, \dots, 40 data for clarity. 
Analogous curves for the scaled spin stiffness, $\rho_{s} L^2$, are shown 
in Fig.~\ref{fig:fssofg}(b), again only for system sizes $L = 30$, 32, \dots, 
40. In both cases, the system sizes are sufficiently large that the crossing 
points exhibit only a very weak dependence on $L$ on the scale used in the 
figure, and both sets of data may be used independently to show that the QCP 
is located at $g_{c} \simeq 4.837(1)$. A detailed analysis is required to 
obtain the most precise results attainable, free of any finite-size effects, 
and we first discuss the general scaling behavior of $g_c$ before presenting
our numerical results.

\subsection{Scaling Forms for Critical-Point Estimators}
\label{sec:qcpscaling}

To describe the evolution of the crossing points with $L$, we perform a 
systematic extrapolation of the finite-size data to the thermodynamic limit 
by extracting the crossing points between data sets for all pairs of system 
sizes, $L$ and $L+2$, based on polynomial interpolations. Figure 
\ref{fig:fssofg}(c) shows the crossing points $g_{c}^R (L)$ and $g_{c}^{\rho} 
(L)$ obtained in this manner.

For any quantity probing a singularity in the thermodynamic limit, one may 
define a size-dependent critical point $g'_{c}(L)$. In general, this quantity 
is expected to shift by an amount proportional to $L^{-1/\nu}$ with respect to 
the true infinite-size critical point, $g_{c}(\infty)$ (hereafter denoted for 
simplicity by $g_{c}$), i.e.~for large $L$,
\begin{equation}
g_{c}(L) = g_{c} + a L^{-1/\nu},
\label{eq:gcpowerform0}
\end{equation}
where $\nu$ is the standard correlation-length exponent. However, with a 
definition based on crossing points of a dimensionless quantity computed 
for two different sizes, the leading corrections cancel and the convergence 
is faster,
\begin{equation}
g_{c}(L) = g_{c} + a L^{-(1/\nu+\omega)}, 
\label{gcpowerform}
\end{equation}
where $\omega > 0$ is the dominant irrelevant exponent. In practice, with
data fits to a rather limited range of available system sizes, the corrections 
to Eq.~(\ref{eq:gcpowerform0}) contained in Eq.~(\ref{gcpowerform}) will have 
exponents and prefactors that deviate from their asymptotic values due to 
the neglected corrections of higher order in $1/L$ and therefore these should 
be considered as ``effective'' quantities. 

The above forms are applicable in the absence of logarithmic corrections, 
but such corrections are the primary focus of our study and are expected at 
$D = D_c$. Kenna has derived the modified form of Eq.~(\ref{eq:gcpowerform0}) 
for classical systems with logarithmic corrections \cite{rkbc,rkb}, 
\begin{equation}
g_{c}(L) = g_{c} + a L^{-1/\nu} \ln^{\hat\lambda} L,
\label{gcpowerformlog0}
\end{equation}
where the exponent of the logarithm for the 4D O$(3)$ universality class is 
$\hat\lambda = - 1/22$. For the crossing points, Eq.~(\ref{gcpowerform}) is
modified to 
\begin{equation}
g_{c}(L) = g_{c} + a L^{-(1/\nu+\omega)} \ln^{\hat c} L,
\label{eq:gcpowerformlog}
\end{equation}
as shown in Appendix A, where $\hat c = \hat\lambda$ if the subleading term 
$L^{-\omega}$ has no multiplicative logarithmic correction, but is altered by 
an unknown amount if it does. Under the circumstances, with a number of 
unknowns and with simulation data only for a restricted range of system 
sizes, we fit our data using not Eq.~(\ref{eq:gcpowerformlog}) but instead 
the purely algebraic form of Eq.~(\ref{gcpowerform}) with $\nu = 1/2$ and 
$\omega$, the effective value of the subleading exponent over the fitting 
range, treated as a different fitting parameter for the separate quantities 
$R_2$ and $\rho_s L^2$. 

\subsection{Numerical Determination of $g_c$}
\label{sec:qcpnumerics}

We take both pairs of fitting parameters $a$ and $\omega$ in 
Eq.~(\ref{gcpowerform}) to be independently free for the two datasets 
$g_{c}^R (L)$ and $g_{c}^{\rho} (L)$, but impose the constraint that the 
curves have the same $g_{c}$. As shown in Figs.~\ref{fig:fssofg}(a) and 
\ref{fig:fssofg}(b), we obtain good fits to both functions, meaning with 
a reduced $\chi^2$ (per degree of freedom, hereafter denoted $\chi_r^2$) 
close to $1$, to the data for all system sizes ($L \ge 6$). These allow 
us to conclude that $g_{c} = 4.83704(6)$, where the numbers in parenthesis 
denote the expected errors (one standard deviation) in the preceding digit, 
i.e.~the relative error on $g_c$ is approximately one part in $10^{5}$. 
If we allow independent parameters $g_c^R$ and $g_c^{\rho}$ in the fits to 
the two data sets, both estimates of the critical point are statistically 
consistent with this $g_c$, albeit with somewhat larger error bars.

We note here that the values we find for the subleading exponent, $\omega 
 = - 0.31(5)$ for the $R_2$ data and $\omega = 0.82(5)$ for the $\rho_s L^2$ 
data, lie far from a common asymptotic value. Thus indeed $\omega$ should be 
considered as an effective exponent accounting for crossover effects in 
system size, neglected higher-order irrelevant fields, and the expected 
weak logarithmic corrections. However, the good match obtained between 
the two extrapolated $g_c$-estimators, especially when approaching the 
infinite-size value from different directions, would not be expected in 
the presence of any corrections not taken sufficiently into account by 
the fitting functions. Thus we believe the error bar on $g_c$ quoted 
above to be completely representative of all statistical and systematic 
uncertainties, in the sense that any remaining systematic errors due to 
the fitting form should be smaller than the statistical errors. The tests 
we perform on the critical scaling behavior around the QCP in the subsequent 
sections also support this statement.

\section{Size-Dependent Logarithmic Corrections at the QCP}
\label{sec:chilogs}

The critical O($N$) $\phi^{4}$ theory, by which is meant the theory at the 
upper critical dimension and at the critical point, obeys many fundamental 
and universal properties, some of which depend on $N$ while others are 
$N$-independent. In Ref.~\cite{Kenna2004} it was shown that the zeros of 
the partition function (Lee-Yang zeros) \cite{YangLee1952}, and hence the 
thermodynamic functions, obey a finite-size scaling theory, which was derived 
by renormalization-group methods. These perturbative arguments demonstrate 
that there are multiplicative logarithmic corrections in the system-size 
dependence of derivable thermodynamic functions, which are closely linked 
to those of the Lee-Yang zeros and, furthermore, are independent of $N$ for
odd $N$. This leads to the key practical observation that size-dependent 
logarithmic corrections in physical observables such as the magnetic 
susceptibility and the specific heat at the critical point follow a 
universal, $N$-independent form when $N = 3$. Here we provide a 
non-perturbative calculation of the magnetic susceptibility, $\chi 
(\mathbf{Q}_{\rm AF},L)$ [Eq.~(\ref{ems})], for systems of finite $L$ at 
the QCP, $g_c$, of the $(3 + 1)$-dimensional O($3$) transition to test 
the predicted logarithmic corrections.

\begin{figure}[t]
\centering
\includegraphics[width=8cm]{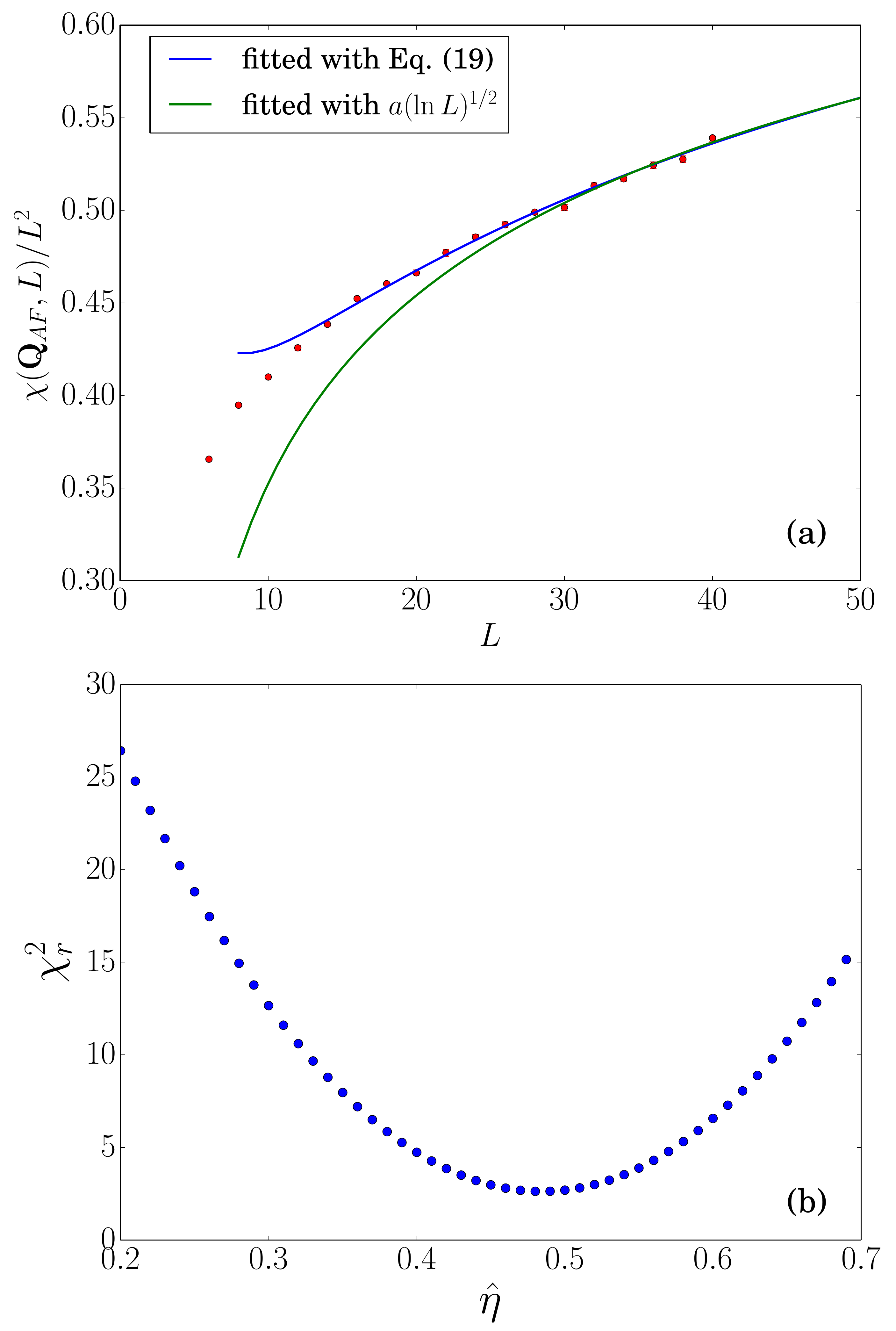}
\caption{(Color online) (a) QMC data for $\chi (\mathbf{Q}_{\rm AF},L) /L^2$ 
obtained at $g = 4.837$ for all even system sizes from $L = 6$ to $40$. 
Solid lines are fits to $a (\ln L)^{1/2}$ (green) and to Eq.~(\ref{eq:chiQAF}) 
(blue). We apply the square-root fit only for system sizes $L \ge 30$ and 
the optimal value of the fitting parameter is $a = 0.274$. The two-parameter 
fit is made to the data for all system sizes $L \ge 14$ and yields optimal 
parameters $a = 0.522$, $b = -1.317$. (b) Reduced $\chi^2$ values obtained 
by fitting $\chi (\mathbf{Q}_{\rm AF},L) /L^2$, for $14 \le L \le 40$, to the 
form (\ref{eq:chiQAF}), but with the exponent $1/2$ replaced by a parameter 
$\hat \eta$. The optimal $\chi_r^2$ value is obtained at $\hat{\eta} \simeq 
0.5$, consistent with the prediction of Ref.~\cite{Kenna2004}.}
\label{fig:chisquare_chiol2}
\end{figure}

The universal form of the magnetic susceptibility at the critical point 
in a finite-size system is given by \cite{Kenna2004} 
\begin{equation}
\chi (\mathbf{Q}_{\rm AF},L) = a L^{2} [\ln L ]^{1/2} \left[ 1 +  
b \frac{\ln(\ln L)}{\ln L} \right] \!\!,
\label{eq:chiQAF}
\end{equation}
with non-universal but $L$-independent parameters $a$ and $b$. We used this 
expression with a fixed value $g = 4.837$, which is within the standard 
deviation of the $g_c$ value found in Sec.~\ref{sec:qcp}, to investigate the 
logarithmic corrections to the $L$-dependence of $\chi (\mathbf{Q}_{\rm AF},L)$. 
We calculate the susceptibility at the ordering wave vector, $\mathbf{Q}_{\rm 
AF} = (\pi,\pi,\pi,\pi)$, for systems of all even sizes from $L = 6$ to 40. 
As in Sec.~\ref{sec:qcp}, the scaling predictions under test are valid for a 
4D theory and again we use the critical contour $T^{-1} = kL$ with $k = 1$.

Figure \ref{fig:chisquare_chiol2}(a) shows our results for the critical 
magnetic susceptibility normalized by $L^{2}$. In the absence of logarithmic 
corrections, $\chi (\mathbf{Q}_{\rm AF},L)/L^2$ would be constant and the curve 
would be a flat line. Instead we observe that a reasonable account of the 
data for our larger system sizes ($L \ge 30$) requires a fit of the form 
$\chi (\mathbf{Q}_{\rm AF},L)/L^2 = a (\ln L)^{1/2}$, as anticipated in 
Ref.~\cite{Kenna2004}. However, for a more quantitative fit over the full 
size range available, we find [Fig.~\ref{fig:chisquare_chiol2}(a)] that it 
is necessary to include the predicted subleading logarithmic correction in 
Eq.~(\ref{eq:chiQAF}). 

To examine the sensitivity of these results to the exponent $1/2$ of the
multiplicative logarithm in Eq.~(\ref{eq:chiQAF}), we replace this predicted 
exponent by a variable ${\hat \eta}$. We determine this exponent by 
calculating the goodness of fit $\chi_r^2$ as a function of $\hat{\eta}$. As 
Fig.~\ref{fig:chisquare_chiol2}(b) makes clear, the best fits are indeed 
obtained close to $\hat{\eta} = 0.5$, in complete consistency with 
Eq.~(\ref{eq:chiQAF}).

From the fact that our exact numerical data confirm not only the leading 
but also the subleading corrections to scaling, we conclude that obvious 
logarithmic corrections can be observed in the size-dependence of the 
thermodynamic functions at the QCP. This result also demonstrates that 
our determination of $g_c$ is sufficiently precise to study logarithmic 
corrections without significant distortions arising from uncertainties in 
its value.

\section{Sublattice Magnetization}
\label{sec:ms}

\begin{figure}[t]
\centering
\includegraphics[width=8.4cm]{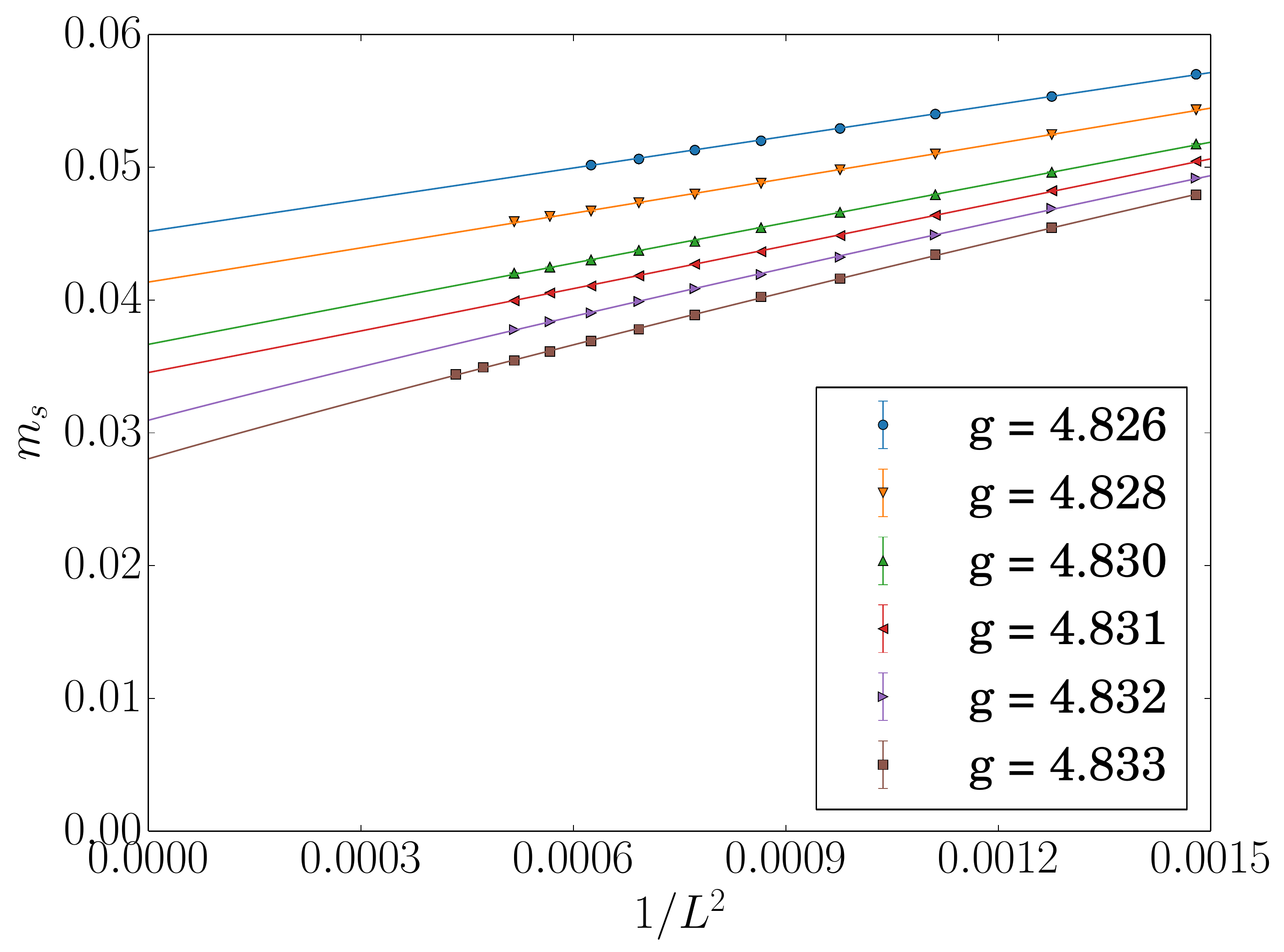}
\caption{(Color online) Staggered magnetization, defined for each system size 
as $m_s(L) = [3 \langle m_{2z}^2(L) \rangle]^{1/2}$, shown as a function of 
$1/L^2$ for a range of coupling ratios $g$ near $g_c$. Polynomial fits of 
cubic order were used to extrapolate $m_s (L)$ to the thermodynamic limit; 
the temperature in all cases was $T = 1/L$. Error bars on all points are 
smaller than the symbol sizes.} 
\label{fig:ms}
\end{figure}

Physical condensed-matter systems at continuous QPTs are generally in the 
thermodynamic limit, and size-scaling measurements of the type easily performed 
in QMC simulations (Sec.~IV) are not a realistic experimental option. However, 
as discussed in Sec.~I, multiplicative logarithmic corrections are expected in 
a range of physical quantities close to the QCP. The primary physical 
observables in the quantum antiferromagnet are the zero-temperature staggered 
magnetization, $m_s (g_c - g)$, and the N\'eel temperature, $T_N (g_c - g)$. 
Calculating these quantities in the thermodynamic limit is significantly more 
challenging than studies of size-dependence, as careful extrapolations of 
finite-size data are required. Here and in Sec.~VI we describe and then  
implement appropriate measures for extrapolating to infinite system size 
and, for $m_s$, to zero temperature, thereby revealing the logarithmic 
corrections to both $m_s$ and $T_N$.

We compute the staggered magnetization according to Eq.~(\ref{ms3mz}) for 
coupling ratios as close to $g_c \simeq 4.837$ as $g = 4.834$. For a given 
value of $g$, we calculate the squared quantity $\langle m^2_{sz}(g,L) \rangle$ 
over a range of system sizes. As shown in Fig.~\ref{fig:ms}, the staggered 
magnetization clearly decreases with increasing $L$ and converges towards a 
fixed limit, suggesting a controlled extrapolation even very close to the 
QCP. Definitive extrapolation to the thermodynamic limit in this regime is 
a complex issue, and a discussion of several technical points is in order 
before analyzing our results. 

\subsection{Extrapolation scheme}

First, most of the simulations in this section are performed at a temperature 
$T = 1/L$, such that the extrapolation $L \rightarrow \infty$ includes both 
system size and temperature. Because the system is ordered for sufficiently 
low $T$ and the order parameter converges quickly to a non-zero value below 
the ordering temperature, one may use the form $T = a L^{-b}$ with arbitrary 
prefactor $a > 0$ and exponent $b > 0$ to study the $T \to 0$ magnetization 
as a function of $L$. These conditions for obtaining a ($T \to 0, L \to 
\infty$) extrapolation of the order parameter for $g < g_c$ contrast with 
the need to follow the contour $T = a L^{-1}$ ($a = 1/k$; $b = 1$ when 
$z = 1$) in Secs.~III and IV for studies of the QCP. Although large values 
of $a$ and $b$ should improve the convergence, in practice one must consider 
the balance between computation time and convergence rate, and the choice 
$a = 1$, $b = 1$ works well in most cases. However, for coupling ratios very 
close to the QCP, the temperature may be a significant fraction of $T_N$ and 
thus $m_s(g,L)$ could be far from its zero-temperature value. We have 
therefore performed additional simulations at $T = 1/(2L)$ to verify that 
the extrapolation does remain well controlled and fully representative of the 
thermodynamic limit in temperature as well as in system size. 

Second, in contrast to Sec.~\ref{sec:qcp}, where we used the non-trivial 
power-law scaling forms (\ref{gcpowerform}) known to be appropriate for 
extrapolating the location of a critical point, the ground-state order 
parameter inside the N\'eel phase can be extrapolated by using simple 
polynomial fits. To obtain $m_s$, one may extrapolate the squared quantity 
and then take its root afterwards or take the square root for each system 
size before extrapolating (the procedure followed in Fig.~\ref{fig:ms}); 
the corresponding polynomials are 
\begin{eqnarray}
m^2_{sz}(g,L)  & = & a(g) + b(g) L^{-2} + c(g) L^{-3} + \dots,  
\label{m2glpolya} \\
\sqrt{m^2_{sz}(g,L)}  & = & a'(g) + b'(g) L^{-2} + c'(g) L^{-3} + \dots 
\label{m2glpolyb} 
\end{eqnarray}
Here the leading $L$-dependence in the extrapolation of a non-vanishing order 
parameter, at fixed $g$ inside the ordered phase, is known \cite{cardy} to be 
$L^{2-d-z}$ due to the dimension-dependent power-law decay of the transverse 
correlation function. Here $d + z = d + 1 = D = D_c = 4$, and the resulting 
leading $1/L^2$ dependence is shown clearly in Fig.~\ref{fig:ms}. Because these 
procedures use a polynomial of finite order to approximate physical behavior 
containing, in principle, an infinite number of corrections, the extrapolated 
value of $m_s$ obtained with the forms (\ref{m2glpolya}) and (\ref{m2glpolyb}) 
will not be exactly the same, but for reliable fits they should agree within 
statistical errors.

We stress that no logarithmic corrections are expected in this case, meaning 
that on grounds of principle they should not be present in the asymptotic 
large-$L$ corrections to a non-zero-valued order parameter. This non-critical 
behavior contrasts with the case of the shift in the critical point discussed 
in Sec.~\ref{sec:qcp}, where logarithmic corrections should in principle be 
present, although we concluded that their effects are not detectable in 
practice. 

Finally, however, non-trivial corrections may still be expected in the $L$ 
dependence of $\langle m^2_{sz}(g,L) \rangle$ for $g$ close to the QCP, where 
the order parameter is small, in the form of crossover behavior from 
near-critical at small system sizes to asymptotic ordered-state scaling at 
large $L$. Quite generally, no analytic functional forms are available for 
describing such crossovers, and great care is required to ensure that the 
asymptotic region, where Eqs.~(\ref{m2glpolya}) and (\ref{m2glpolyb}) are 
valid, has been reached. As $g \rightarrow g_c$, successively larger systems 
are required for this, and here we find that reliable extrapolations are no 
longer possible beyond $g = 4.834$ because of the limits on system size set 
by the available computer resources.

\begin{figure}[t]
\centering
\includegraphics[width=8.0cm]{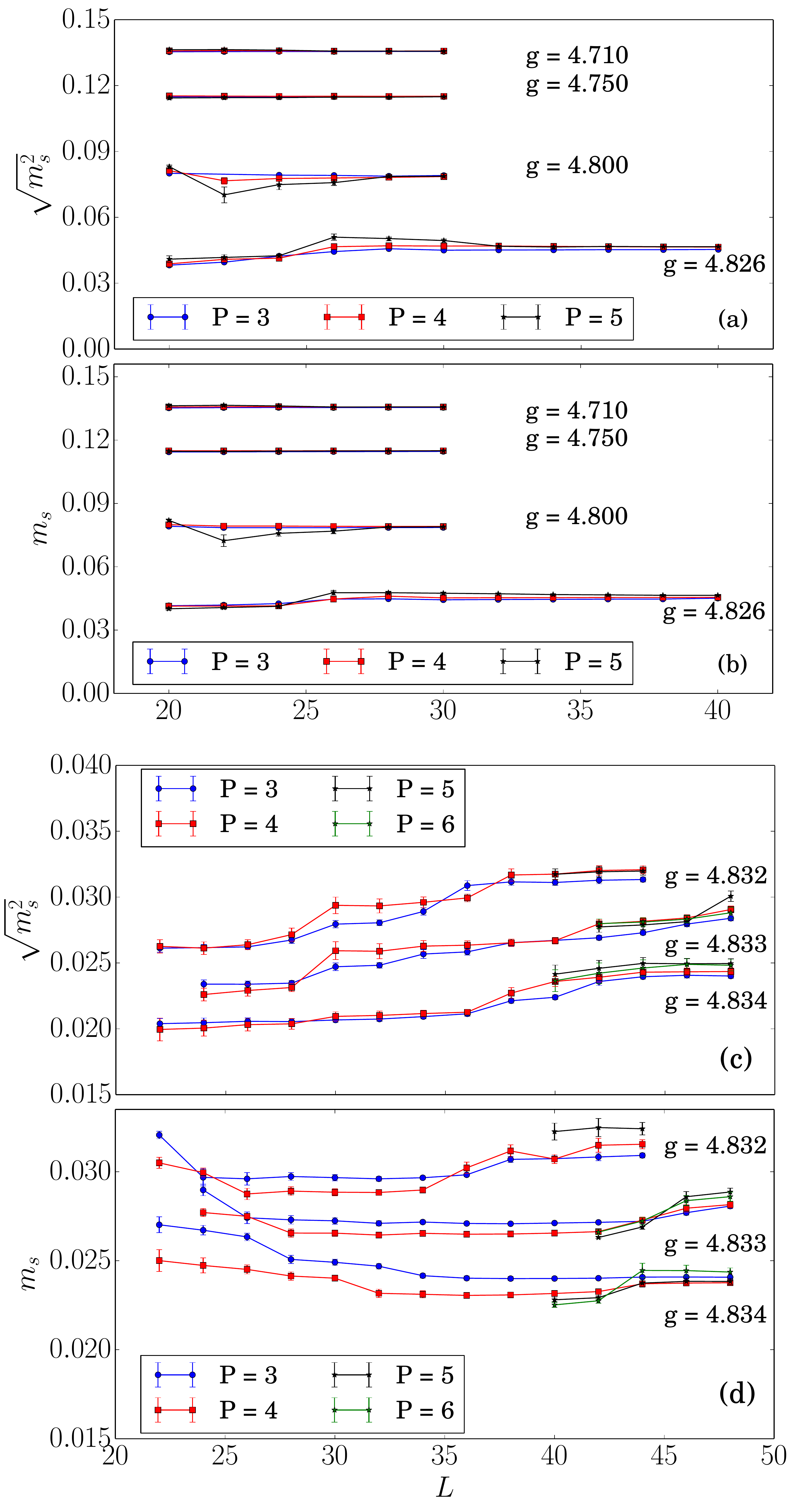}
\caption{(Color online) Extrapolated values of the sublattice magnetization 
as a function of the largest system size included in the fit, for values $g$ 
near the QCP ($g_c \simeq 4.837$). Panels (a) and (b) include values of $g$ 
for which the extrapolations are stable for relatively small $L$; panels (c) 
and (d) show $g$ values very close to $g_c$, where the extrapolations require 
large sizes to stabilize. Results for different orders $P$ of the fitting 
polynomial are compared. The smallest system size included was determined 
using the $\chi_r^2$ criterion of Eq.~(\ref{chi2crit}). Panels (a,c) 
and (b,d) show respectively the results of extrapolations of $\langle m^2_{sz} 
(g,L) \rangle$ and $\langle m^2_{sz}(g,L) \rangle^{1/2}$, with the square root 
taken after the extrapolation in the former case.}
\label{fig:ms2}
\end{figure}

In fits to the forms (\ref{m2glpolya}) and (\ref{m2glpolyb}), it is necessary 
to select the order $P$ of the polynomial and the range of system sizes to 
include. The size of the error bars on the QMC data points has a significant 
influence here, because deviations from the leading $L^{-2}$ correction 
are easier to detect with smaller error bars. We have performed a systematic 
study using fits of orders $P = 3$ to 6, including different system-size 
ranges. We characterize the quality of the fits using the standard reduced 
$\chi^2$ measure, and for a ``good'' fit we require that the optimal value 
must fall within three standard deviations of its mean, i.e.~we demand that
\begin{equation}
\chi_r^2 - 1 = \frac{\chi^2}{n_L - n_p} - 1 \le 3 \sqrt{\frac{2}{n_L
 - n_p}}, 
\label{chi2crit}
\end{equation}
where $n_L$ is the number of data points (system sizes) and $n_p = P + 1$ is 
the number of fitting parameters. For a given $P$ and largest system size $L$, 
we use all available system sizes down to a smallest size $L_{\rm min}$ for 
which the above condition is still satisfied. We then study the behavior as a 
function of $L$ for different $P$, and compare the values of $m_s$ obtained 
from extrapolations based on Eqs.~(\ref{m2glpolya}) and (\ref{m2glpolyb}).
To estimate the error bars on the extrapolated $m_{s}(g)$, we performed 
additional polynomial fits with Gaussian noise (whose standard deviation is 
equal to the corresponding QMC error bars) added to the finite-size data. 
The standard deviation of the distribution of extrapolated $m_{s}(g)$ values 
defines the statistical error. 

Figure~\ref{fig:ms2} shows results for several values of $g$ approaching the 
QCP, with cases where the fits are relatively straightforward (further from 
$g_c$) shown in panels (a) and (b) and more challenging cases (closer to $g_c$) 
shown in panels (c) and (d). The upper panel in each group corresponds to the 
square root being taken after the extrapolation [Eq.~(\ref{m2glpolya})], while 
the lower corresponds to fitting the square root for each system size 
[Eq.~(\ref{m2glpolyb})]. In panels (a) and (b), the extrapolated values are 
observed to be very stable with respect to the range of system sizes and the 
order of the polynomial, whereas panels (c) and (d) manifest some of the 
crossover behavior expected close to $g_c$, showing considerable variation 
as the maximum system size is increased. There are also significant differences 
between the two fitting procedures, until the largest system sizes where the 
extrapolations stabilize; we take the fact that the two types of fits give 
consistent results for these largest systems at all values of $g$ as an 
indication that the extrapolations are reliable. We have not been able to 
achieve good convergence based on system sizes up to $L = 48$ for $g$ values 
closer to $g_c$ than those shown in Figs.~\ref{fig:ms2}(c) and (d). In these 
most challenging cases, our results show that it is better to use the fitting 
form of Eq.~(\ref{m2glpolyb}), extrapolating the square root of the staggered 
magnetization for each system. Regarding the quality of the fits obtained by 
varying the polynomial order, $P$, in Fig.~\ref{fig:ms2}, we find that extra 
terms in the fit scarcely justify the additional degrees of freedom lost in 
the determination of $\chi_r^2$. All of the results presented below were 
obtained by extrapolating $\sqrt{m_s^2}$ with polynomials of order $P = 4$.

\begin{figure}[t]
\includegraphics[width=\columnwidth]{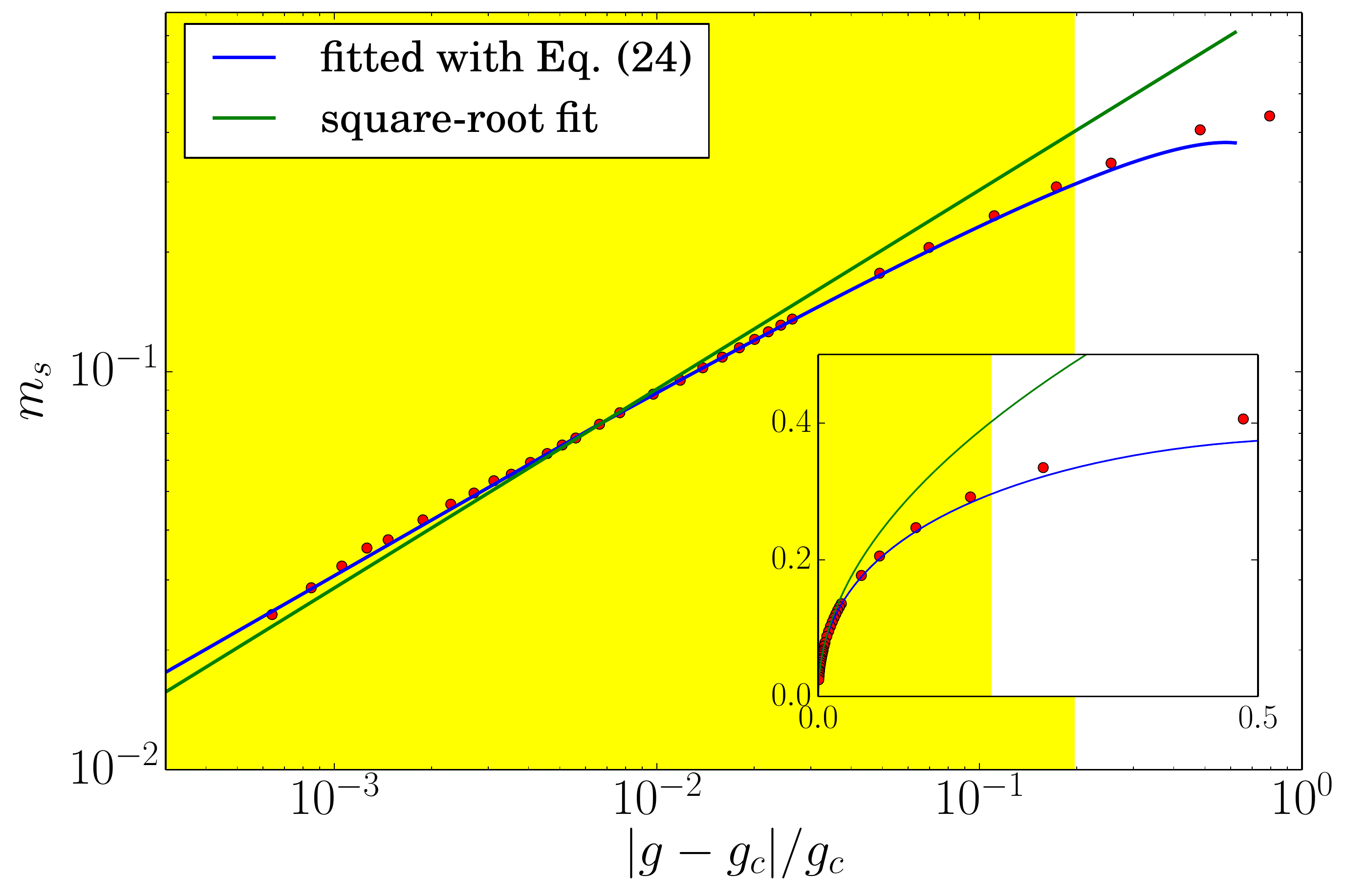}
\caption{(Color online) Extrapolated staggered magnetization, $m_s$ at $T = 0$,
as a function of the distance from the QCP, using the value $g_c = 4.83704$ 
determined in Sec.~\ref{sec:qcp}. Error bars on the calculated data points 
are similar to or smaller than the symbol size. The closest point to $g_c$ is 
$g = 4.834$. Lines show both the best fit by a pure square-root function 
[Eq.~(\ref{eq:sqrt}), green] and including the logarithmic correction factor 
predicted in Ref.~\cite{Kenna2004} [Eq.~(\ref{eq:log}), blue]. The fitting 
parameters of the logarithmic correction curve are $a = 0.266(2)$ and $b = 
4.8(3)$. The yellow shading represents the approximate extent of the QC 
regime and is determined by including all data points described adequately 
(within a deviation of approximately $4\%$, see text) by the functional form 
of the logarithmic correction curve. The inset shows $m_s(g)$ and the QC 
regime on linear axes.}
\label{fig:msg}
\end{figure}

\subsection{Thermodynamic Limit}

With all of the above considerations, we are able to obtain reliable and 
high-precision extrapolations of the staggered magnetization in the 
thermodynamic limit for values of $g$ as close to the QCP as $|g - g_c| 
\simeq 0.003$. In Fig.~\ref{fig:msg} we show all of our data for $m_s 
(|g - g_c|)$ on logarithmic axes. If these data satisfied mean-field scaling 
alone, with no discernible logarithmic corrections, one would expect a curve 
of the form 
\begin{equation}\label{eq:sqrt}
m_s(g) = a |g - g_c|^{1/2},
\end{equation}
but this (green line in Fig.~\ref{fig:msg}) is manifestly unable to 
describe the data. For the zero-temperature order parameter, perturbative 
renormalization-group considerations applied to the O($N$) $\phi^4$ field 
theory at the upper critical dimension predict the form 
\begin{equation}\label{eq:log}
m_s(g) = a |g - g_c|^{\beta} |\ln(|g - g_c|/b)|^{\hat{\beta}},
\label{mslogformbeta}
\end{equation}
where $\beta = 1/2$ is the mean-field exponent and the exponent of the 
multiplicative logarithmic correction is given by $\hat{\beta} = 3/(N + 8)$ 
\cite{Kenna2004}. A fit to this form, using $\hat{\beta} = 3/11$ for $N = 3$ 
(blue curve in Fig.~\ref{fig:msg}), yields excellent agreement with the data 
all the way to our smallest values of $|g - g_c|$; the fitting parameters 
are $a = 0.266 \pm 0.002$ and $b = 4.8 \pm 0.3$. We note that the fit is very 
insensitive to the precise value of $b$, and for further analysis we fix this
to $b = g_c$.

\begin{figure}[t]
\includegraphics[width=\columnwidth]{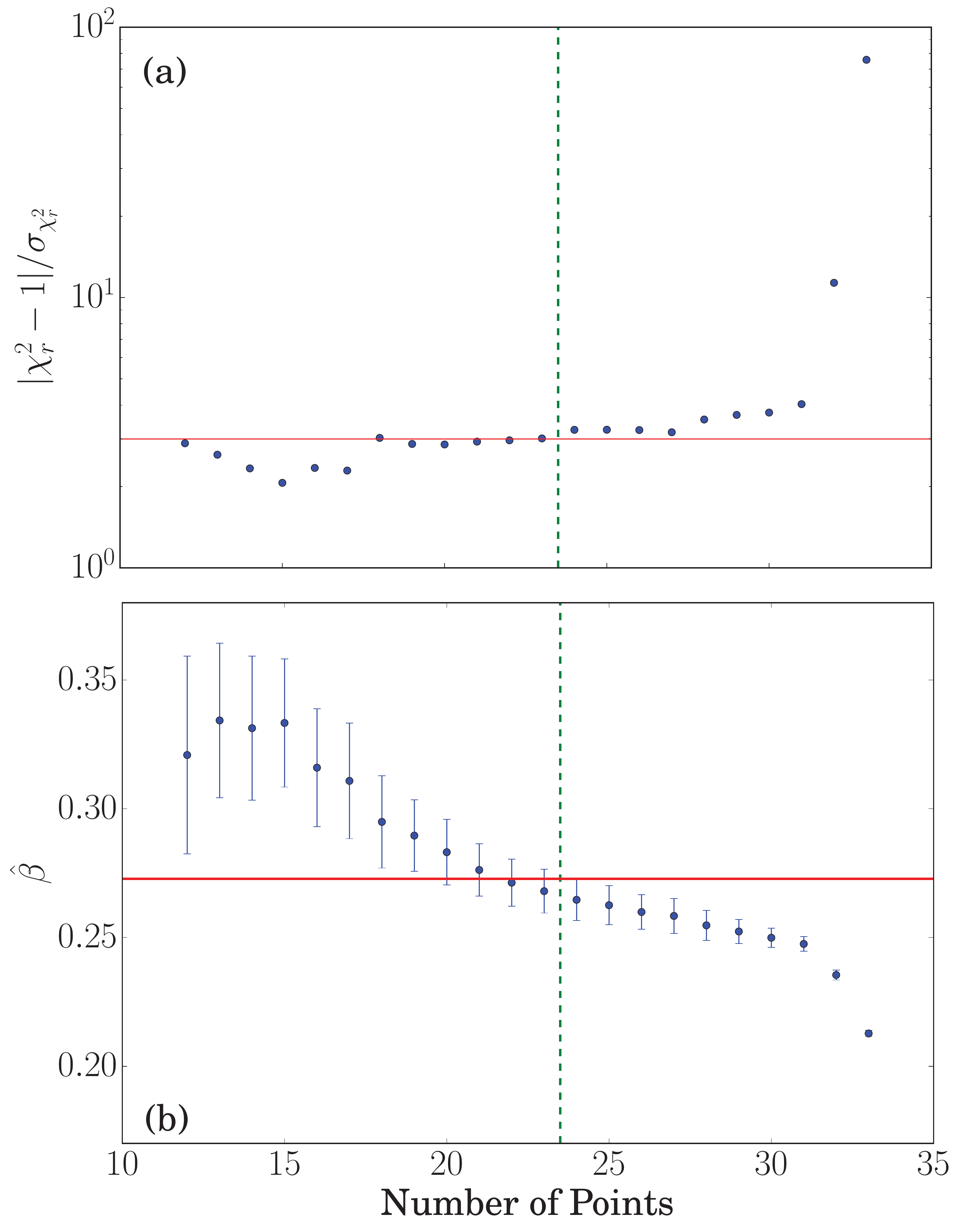}
\caption{(Color online) Statistical analysis of the exponent of the 
logarithmic correction in Eq.~(\ref{mslogformbeta}). (a) Reduced $\chi^2$ 
value of the fit, normalized to the standard deviation, and (b) optimal 
value of the exponent shown as functions of the number of data points 
($g$-values) used, beginning from the point closest to $g_c$ in 
Fig.~\ref{fig:msg}. The vertical dashed line indicates the number of 
points, $N_g = 23$, for which a $3\sigma$ criterion for $\chi_r^2$ 
[cf.~Eq.~(\ref{chi2crit})] is satisfied, as indicated by the horizontal 
line in panel (a). In panel (b), the error bars were computed by repeating 
the fits multiple times with Gaussian noise added to the $m_s$ data points. 
The horizontal line marks the predicted value $\hat{\beta} = 3/11$.}
\label{fig:ms-log}
\end{figure}

To test the predicted exponent $\hat\beta = 3/11$ in Eq.~(\ref{mslogformbeta}), 
we treat it as a free parameter and fit our data using different numbers of 
$g$ values, including all points closest to $g_c$ and studying the behavior 
as points further away from $g_c$ are added one by one. Figure \ref{fig:ms-log} 
shows $\chi_r^2$ and $\hat\beta$ as functions of the number of data 
points fitted. With the exception of cases including the two points furthest 
away from $g_c$, all the fits appear reasonable, with $\chi_r^2 < 2$.  
However, by the properties of the $\chi^2$ distribution, a fit should be 
considered statistically acceptable only if a criterion analogous to 
Eq.~(\ref{chi2crit}) is satisfied, i.e.~the largest number of data points 
for which $\chi_r^2 - 1$ remains less than three times its standard deviation 
($3 \sigma$) marks the boundary between good and poor fits. At this point we 
obtain $\hat{\beta} = 0.268 \pm 0.008$, which lies well within one standard 
deviation of the predicted value $3/11 \approx 0.2727$. If more points are 
excluded, the fitted exponent evolves slowly [Fig.~\ref{fig:ms-log}(b)] 
while remaining statistically well compatible with the predicted value. 
Because the fitting error increases, less weight should be placed on results 
including less data, and taking an error-weighted average over all the points 
below the cut-off line, $N_g = 23$, in Fig.~\ref{fig:ms-log} yields 
$\hat\beta = 0.279 \pm 0.011$. We take this as complete confirmation 
of the predicted value.

As important as finding clear logarithmic corrections to scaling is that we 
have demonstrated their presence over a significant region around the QCP; 
indeed, most of the points we have computed are well described by 
Eq.~(\ref{eq:log}). Including the multiplicative logarithmic correction 
converts an inadequate description of the data into an excellent one 
(Fig.~\ref{fig:msg}) as far inside the N\'eel phase as $|g - g_c|/g_c 
\approx 0.2$, where the order parameter is already at $60\%$ of its maximum 
possible value ($m_s = 1/2$, at which point no quantum fluctuation effects 
remain). This improvement is clearer still in the inset of Fig.~\ref{fig:msg}, 
which shows the results on linear axes. Under the assumption that data points 
at large $|g - g_c|$ no longer fall on the fitted curve because they lie 
outside the region controlled by the QCP, we can determine the size of the 
critical region based on a threshold maximum deviation of the data from the 
curve. Although the choice of threshold value is somewhat arbitrary, the 
$|g - g_c|/g_c \le 0.2$ region indicated by the yellow shading in 
Fig.~\ref{fig:ms} reflects a threshold of approximately $4\%$, which 
lies well above achievable experimental uncertainties. We comment in 
Sec.~\ref{sec:summary} on the utility of our results for the case of 
TlCuCl$_3$. 

\section{N\'eel temperature}
\label{sec:tn}

We turn next to the scaling form of the N\'eel temperature, $T_N(g)$, near 
the QCP. Unlike the $T = 0$ order parameter, as far as we are aware there 
is no prediction from perturbative field-theoretical calculations including 
logarithmic corrections for the scaling form of finite-$T$ critical points 
at the upper critical dimension. Close to a QCP, the general power-law form 
without logarithmic corrections is discussed in Ref.~\cite{rs}, but one 
may also expect a multiplicative logarithmic term as in the other quantities 
we have discussed. We first derive the exponent of the logarithm for the 
O($3$) transition in 3+1 dimensions, based on the known scaling properties 
of related quantities. We then present our QMC calculations of $T_N$ for the 
Heisenberg model on the double cubic lattice and test our prediction.

\subsection{Scaling hypothesis}
\label{subsec:3}

In a path-integral construction in imaginary time, the size of the system 
in the time dimension is proportional to the inverse temperature, $\beta = 
1/T$. This can be considered as a length, $L_\tau$, on a parallel with the 
spatial lengths, $L$, of a $d+1$-dimensional system. If the spatial lengths 
are taken to infinity in all directions, what remains is a single finite 
length, $L_\tau$, for the effective system, and finite-size scaling in this 
length corresponds to finite-$T$ scaling in the original quantum system 
\cite{Chakravarty88}.

Without logarithmic corrections, by analogy with the finite-$L$ shift of 
the critical point discussed in Sec.~\ref{sec:qcpscaling}, the same type of 
shift as in Eq.~(\ref{eq:gcpowerform0}) can be expected because $z = 1$. Thus 
\begin{equation}
g_c(T) - g_c(0) \sim L_\tau^{-1/\nu}, 
\end{equation}
as a consequence of the finite temporal size, and the scaling behavior 
is $T_N \sim (g_c - g)^\nu$, as discussed in detail in Ref.~\cite{rs}. In
the case of spatial finite-size scaling, with all lengths finite, the shifted 
critical point (sometimes called the pseudo-critical point) is not a singular 
point, but the singularity develops as $L \to \infty$. By contrast, in the 
finite-$T$ case in $d = 3$ spatial dimensions, the shifted point is a true 
(classical) phase transition, although from a scaling perspective this 
difference is not relevant.

In order to discuss logarithmic corrections, it is useful to first express 
$T_N$ using a macroscopic, zero-temperature energy scale of the system that 
vanishes as $g \to g_c$ \cite{rs}. For the spin system considered here, the 
only such energy scale is the spin stiffness, $\rho_s$. According to 
Ref.~\cite{Fisher1989}, the scaling form of this quantity in the ordered 
phase when $z = 1$ is
\begin{equation}
\rho_s \sim (g_c - g)^{\nu(d-1)}.
\end{equation}
Consistency with the result $T_N \sim (g_c - g)^\nu$ then gives the scaling 
of the critical temperature for $d = 3$,
\begin{equation}
T_N^2 \sim \rho_s,
\label{eq:tnrhos}
\end{equation}
where the mismatch in units is compensated by a power of the non-singular 
spin-wave velocity \cite{rs} (Sec.~VIC), which can be neglected here. Our 
basic hypothesis is that this proportionality, which is the singular part 
of a relationship based on matching scaling dimensions, applies in all 
respects at the upper spatial critical dimension ($d = 3$ for $z = 1$), 
such that logarithmic corrections to $T_N$ arise solely due to the 
logarithmic corrections intrinsic to $\rho_s$.

Fisher {\it et al.}~\cite{Fisher1989} have shown that the critical spin 
stiffness can be expressed as $\rho_s \sim \xi^2 f$, where $\xi$ is the 
correlation length and $f$ is the free-energy density. The logarithmic 
corrections to both $\xi$ and $f$, presented by Kenna in Ref.~\cite{rkbc}, 
are 
\begin{equation}
\xi \sim |g - g_c|^{-\nu} \ln^{\hat\nu} (|g - g_c|),
\end{equation}
with $\hat\nu = 5/22$ for the relevant universality class, and 
\begin{equation}
f \sim |g - g_c|^{4\nu} \ln^{\hat\alpha} (|g - g_c|), 
\end{equation}
with $\hat\alpha = 1/11$. The logarithmic correction to $\rho_s$ is 
therefore given by
\begin{equation}
\rho_s \sim |g - g_c|^{2\hat\nu} \ln^{2\hat\nu + \hat\alpha} (|g - g_c|),
\end{equation}
and by combining these results with Eq.~(\ref{eq:tnrhos}) we obtain
\begin{equation}
T_N \sim |g - g_c|^{\nu} \ln^{\hat\tau} (|g - g_c|),
\label{tnlog322}
\end{equation}
where $\hat\tau = \hat\nu + {\hat\alpha}/{2}$. From the values of $\hat\nu$ 
and $\hat\alpha$ given above \cite{rkbc}, we obtain the prediction $\hat\tau
 = 3/11$, which is remarkable in that the exponent in the logarithmic 
correction to $T_N$ should be the same as the one in the sublattice 
magnetization, $\hat\tau = \hat\beta$ (\ref{mslogformbeta}). 

Because the zero-temperature order parameter is a consequence purely of 
quantum fluctuations, whereas the classical ordering temperature is a 
consequence primarily of thermal fluctuations, there is {\it a priori} 
no reason to expect that the two should have the same form. Exact numerical 
calculations are therefore uniquely positioned to provide qualitatively new 
information in this case. We note that this equality applies to the phase 
transitions of O($N$) models for all values of $N$; because $\hat\nu = 
(N+2)/[2(N+8)]$ and $\hat\alpha = (4-N)/(N+8)$ \cite{rkbc}, we obtain
$\hat\tau = 3/(N+8)$, the same value as the exponent $\hat\beta$ in 
Eq.~(\ref{mslogformbeta}). Thus we predict that, in the neighborhood of 
$g_c$, $T_N(g)$ will be proportional to $m_s(g,T = 0)$, with no multiplicative 
logarithmic factors, for all values of $N$; this result was reported for the 
$N = 3$ case in a previous QMC study \cite{Jin2012}, which we now extend 
sufficiently close to the QCP to observe the cancellation of logarithmic 
terms.

\subsection{QMC calculations}
\label{subsec:4}

Calculating $T_N(g)$ within our QMC simulations is similar to obtaining $g_c$ 
in Sec.~\ref{sec:qcp}, but with some important differences of detail. The 
calculations of Sec.~\ref{sec:qcp} were performed for a genuinely 4D system, 
with the imaginary-time axis treated on the same footing as the spatial 
dimensions. At finite temperatures, this symmetry is broken and the system 
is 3D with a separate temperature variable, which determines the finite 
thickness of the time dimension even when $L \to \infty$. Both the Binder 
ratio [Eq.~(\ref{er2})] and the spin stiffness [Eq.~(\ref{ess})] are 
size-independent quantities at the thermal phase transition and hence remain 
valuable indicators, although the appropriately scaled spin stiffness for the 
3D transition is now $\rho_{s}L$ (instead of $\rho_{s}L^2$, used for analyzing 
the 4D $T = 0$ transition in Sec.~\ref{sec:qcp}) \cite{Jin2012}.

\begin{figure}[t]
\centering
\includegraphics[width=\columnwidth]{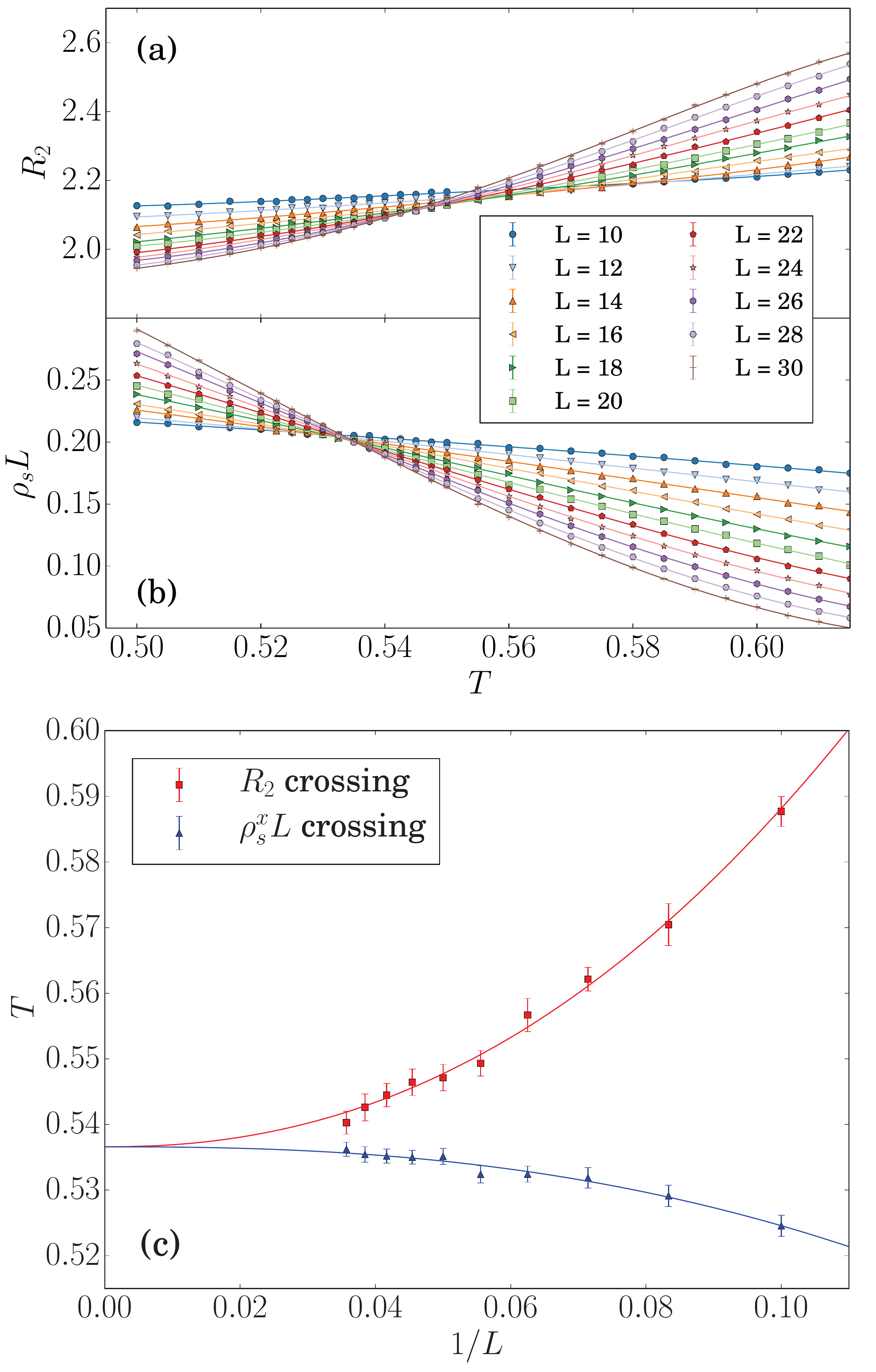}
\caption{(Color online) Procedures used to extract the N\'eel temperature, 
$T_N(g)$, illustrated for the case $g = 4.71$. (a) Binder ratio $R_{2}$ as a 
function of $T$ for system sizes $L = 10$, 12, \dots, 30. (b) Scaled spin 
stiffness $\rho_{s} L$ as a function of $T$ for the same values of $L$. 
Error bars are smaller than the symbol sizes. Crossings of these lines are 
extracted using polynomial fits and the results are used to obtain finite-size 
estimates for quantities $T_N^R (L)$ and $T_N^{\rho} (L)$. (c) Fits to data for 
systems of all sizes ($L \ge 6$) of the two size-dependent crossing estimators 
using functions of the form $T_N (L) = T_N (\infty) + a/L^{1/\nu+\omega}$. 
Enforcing the same constant $T_N (\infty) \equiv T_N$ for the $R_2$ and 
$\rho_{s} L$ crossings gives $T_N = 0.5363(13)$ with irrelevant exponents 
$\omega \approx 0.8(2)$ for $R_2$ and 1.1(3) for $\rho_{s} L$ ($1/\nu \approx 
1.42$ for the relevant 3D universality class).}
\label{fig:tncrossing}
\end{figure}

For each value of the coupling ratio $g$ within the N\'eel phase ($g < 
g_{c}$), we compute $R_{2}$ and $\rho_{s}L$ for a range of system sizes 
and perform finite-size-scaling extrapolations to deduce the N\'eel 
temperature, $T_N(g)$, in the thermodynamic limit. Similar to 
Sec.~\ref{sec:qcp}, we first obtain the crossings of the $R_2(T)$ and 
$\rho_{s}L (T)$ data for different system sizes using polynomial 
fits, as shown in Figs.~\ref{fig:tncrossing}(a) and \ref{fig:tncrossing}(b) 
for $g = 4.71$. The crossing points of both quantities for each successive 
pair of system sizes, $T_N(g,L)$ and $T_N(g,L+2)$, are used to extrapolate 
towards the value $T_N(g,L \to \infty)$ from above and below, using power-law 
forms analogous to Eq.~(\ref{gcpowerform}). We note that, as in the analysis 
leading to $g_c$ (Sec.~\ref{sec:qcpnumerics}), the data points obtained for 
$R_2(L,T)$ and $\rho_{s}L (L,T)$ using systems of all sizes ($L \ge 10$) fall 
within a $3 \sigma$ criterion analogous to Eq.~(\ref{chi2crit}) for this 
type of fit. The extrapolation of $T_N (g, L \to \infty) \equiv T_N(g) = 
0.5363(13)$ for $g = 4.71$ is shown in Fig.~\ref{fig:tncrossing}(c).

We comment here that our determination of $T_N(g)$ for $g$ close to $g_c$ 
is rather less precise than our determination of $m_s(g)$. The fundamental 
difference in character of the two quantities, and hence of their 
calculation, causes the estimators for $T_N(g)$ [the approximate crossings 
in Figs.~\ref{fig:tncrossing}(a) and \ref{fig:tncrossing}(b)] to have larger 
error bars and finite-size effects. Further, the error bars of the crossing 
points grow rapidly as $g \to g_c$, while the decrease in $T_N$ leads to 
longer simulation times (because the space-time volume is proportional to 
$L^3/T$). After detailed error control, the closest reliable data point to 
the QCP is $g = 4.831$, for which $|g - g_c|$ is twice as large as for the 
closest $m_s(g)$ point (Sec.~\ref{sec:ms}). We have nevertheless obtained 
19 reliable data points, within the QC regime determined from $m_s(g)$ 
(Fig.~\ref{fig:msg}) and down to unprecedentedly low temperatures, which 
are fully sufficient to test for evidence of logarithmic corrections to 
$T_N(g)$. 

\begin{figure}[t]
\includegraphics[width=\columnwidth]{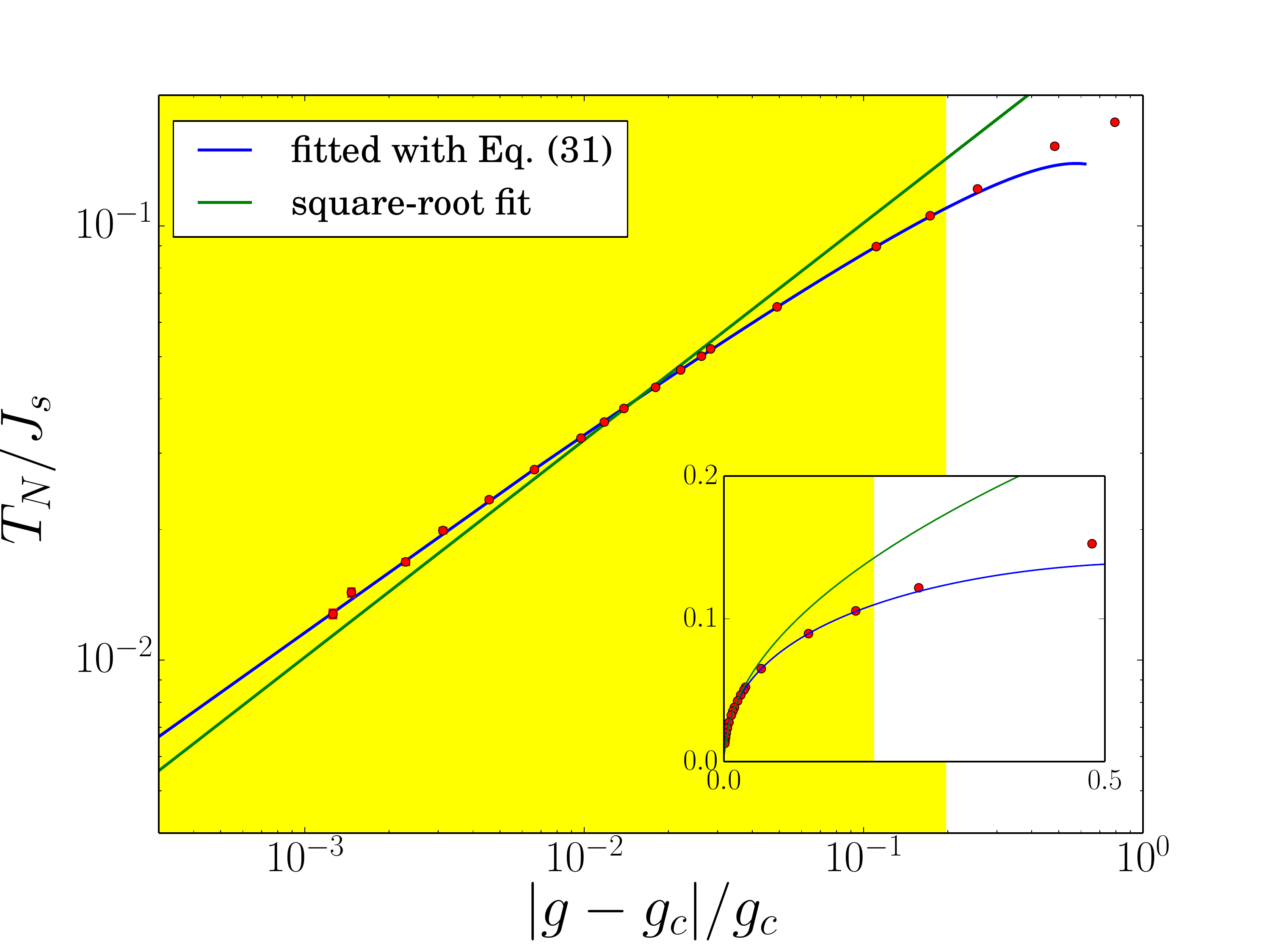}
\caption{(Color online) Normalized N\'eel temperature, $T_N/J_s$, as a 
function of the distance from criticality, $|g - g_c|$. The closest point 
to $g_c \approx 4.837$ is $g = 4.831$. Lines show both the best fit by a 
pure square-root function (green) and using a logarithmic correction factor 
with exponent $\hat\tau = 3/11$ [Eq.~(\ref{tnlog322}), blue]. The yellow 
shading represents the QC regime and is determined from Fig.~\ref{fig:ms2}. 
The inset shows $T_N(g)$ on linear axes.}
\label{fig:tng}
\end{figure}

The N\'eel temperature has units of energy and clearly depends on the 
overall energy scale of the system. Ideally, it should be normalized by an 
intrinsic energy scale of the system to give a dimensionless quantity. In 
Ref.~\cite{Jin2012} it was shown that $T_{N} (|g - g_c|)$ normalized to the 
microscopic energy scale $J_s$, given by the sum of all couplings of a 
spin to its neighbors, yields a remarkably system-independent result for 
Heisenberg antiferromagnets with three different dimerization patterns;
for the double cubic lattice, $J_s = J(6 + g)$. Other authors 
\cite{Oitmaa2012,Kao2013} have suggested that the appropriate normalization 
is given by a macroscopic quantity, the spatially averaged spin-wave velocity, 
$\sqrt{c_x c_y c_z}$ (which, it should be noted, does not have units of energy
and requires an unknown dimensionful constant). We begin by taking the former 
approach and return below to address the latter. 

\begin{figure}[t]
\includegraphics[width=\columnwidth]{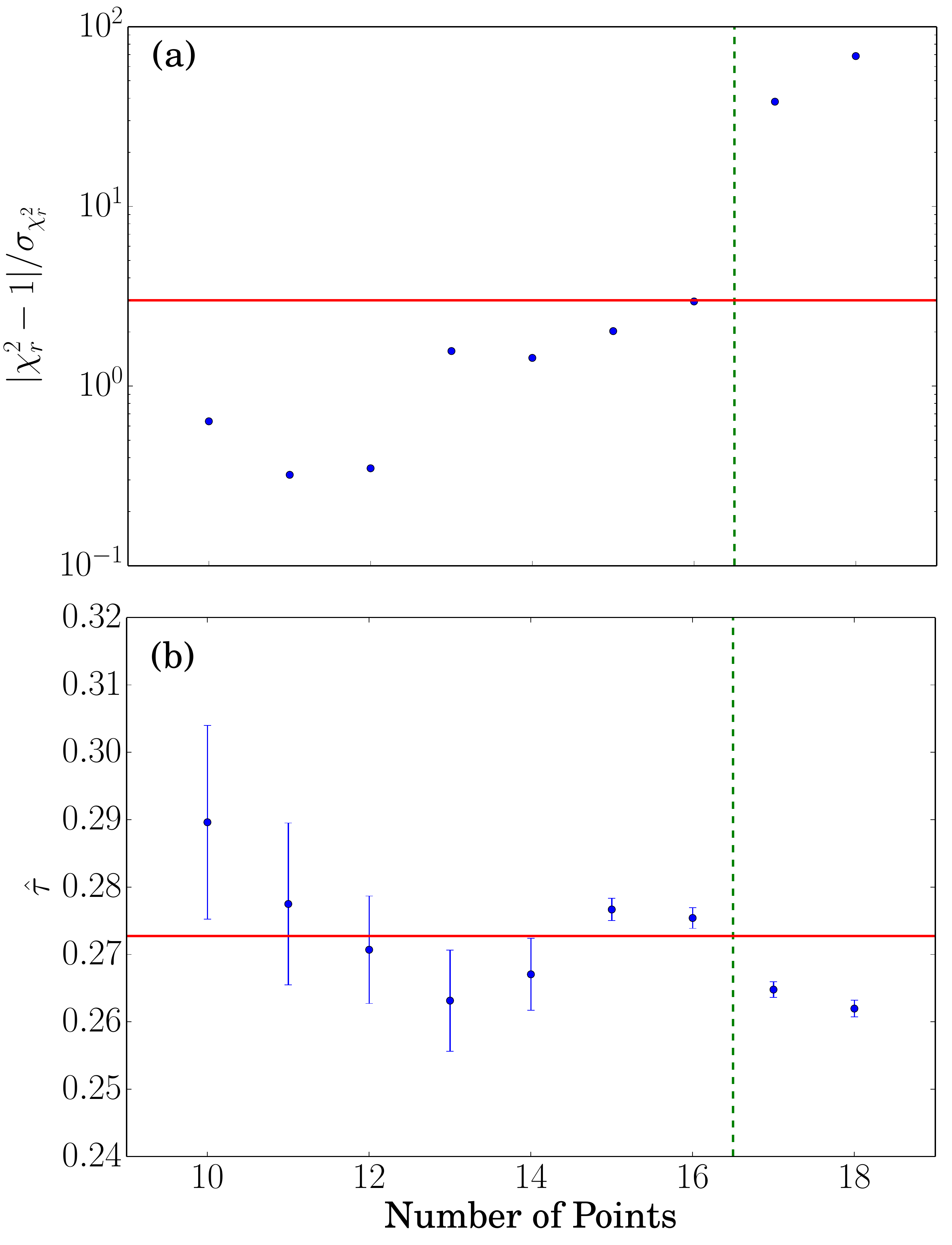}
\caption{(Color online) Statistical analysis of the exponent of the 
logarithmic correction in Eq.~(\ref{tnlog322}), performed by replacing the 
predicted value $3/11$ with a fitting parameter $\hat \tau$. (a) Reduced 
$\chi^2$ value of the fit, normalized to the standard deviation, and (b) 
optimal value of the exponent, both shown as functions of the number of data 
points ($g$-values) used, beginning from the point closest to $g_c$ in 
Fig.~\ref{fig:tng}. The vertical dashed line indicates the number of 
points, $N_g = 16$, included in the fit below which $\chi_r^2$ 
satisfies a $3\sigma$ criterion analogous to Eq.~(\ref{chi2crit}), as 
indicated by the horizontal line in panel (a). In panel (b), the error 
bars were computed by repeating the fits multiple times with Gaussian 
noise added to the $T_N$ data points. The horizontal line marks the 
predicted value $\hat\tau = 3/11$.}
\label{fig:tn-log}
\end{figure}

The relationship between $T_N/J_s$ and $|g - g_c|$ is presented in 
Fig.~\ref{fig:tng}. Once again we show a mean-field scaling line for 
comparison and once again it cannot provide an adequate fit, suggesting 
that logarithmic corrections are indeed present. However, a fit to our 
predicted form, given by Eq.~(\ref{tnlog322}) with $\hat\tau = 3/11$, 
describes the data very well, even at the limits of the region classified 
as QC based on the $m_s(g_c - g)$ fit in Fig.~\ref{fig:ms2}.

For a fully quantitative test of the exponent we predict for the multiplicative
logarithmic term in Eq.~(\ref{tnlog322}), we substitute a free exponent $\hat 
\tau$ for the fixed value $3/11$ and optimize it using fits with different 
windows of $g$-values. This analysis is precisely analogous to that performed 
for $m_s$ in Fig.~\ref{fig:ms-log}. The behavior of $\chi_r^2$ and of 
the optimized exponent, with error bars again estimated using the method of 
numerical Gaussian noise propagation, is presented in Fig.~\ref{fig:tn-log}. 
By taking the inverse-variance-weighted average over all results for which 
$\chi_r^2$ is acceptable, we obtain $\hat\tau = 0.275(2)$, in excellent 
agreement with the prediction $\hat\tau = 3/11 \simeq 0.2727$. We conclude 
that the multiplicative logarithmic correction to $T_N(g)$ is, to within our 
error bars and in agreement with a straightforward scaling argument based on 
the spin stiffness (Sec.~VIA), identical to the $m_s(g)$ correction in 
Eq.~(\ref{eq:log}).

\subsection{Spin-wave velocity}
\label{subsec:5}

The spin-wave velocity, $c$, is uniform in the primary axial directions on 
the double cubic lattice. As discussed in Sec.~\ref{sec:methods}, it can be 
calculated most straightforwardly and most accurately in the SSE framework from
the spatial and temporal winding-number fluctuations in Eq.~(\ref{winding}) to 
define the space-time-isotropic criterion of Eq.~(\ref{spinwavevelocity}), 
which contains the velocity \cite{Kaul2008,Jiang2011,Kao2013,Sen2015}. This 
technique remains well-defined throughout the critical regime and is 
expected not to be affected by logarithmic corrections; it was shown in 
Ref.~\cite{Sen2015}, which we follow for technical details, that the 
winding-number approach produces the correct result for $c$ in the Heisenberg 
chain, a system known to have strong logarithmic corrections to scaling. 

\begin{figure}[t]
\includegraphics[width=\columnwidth]{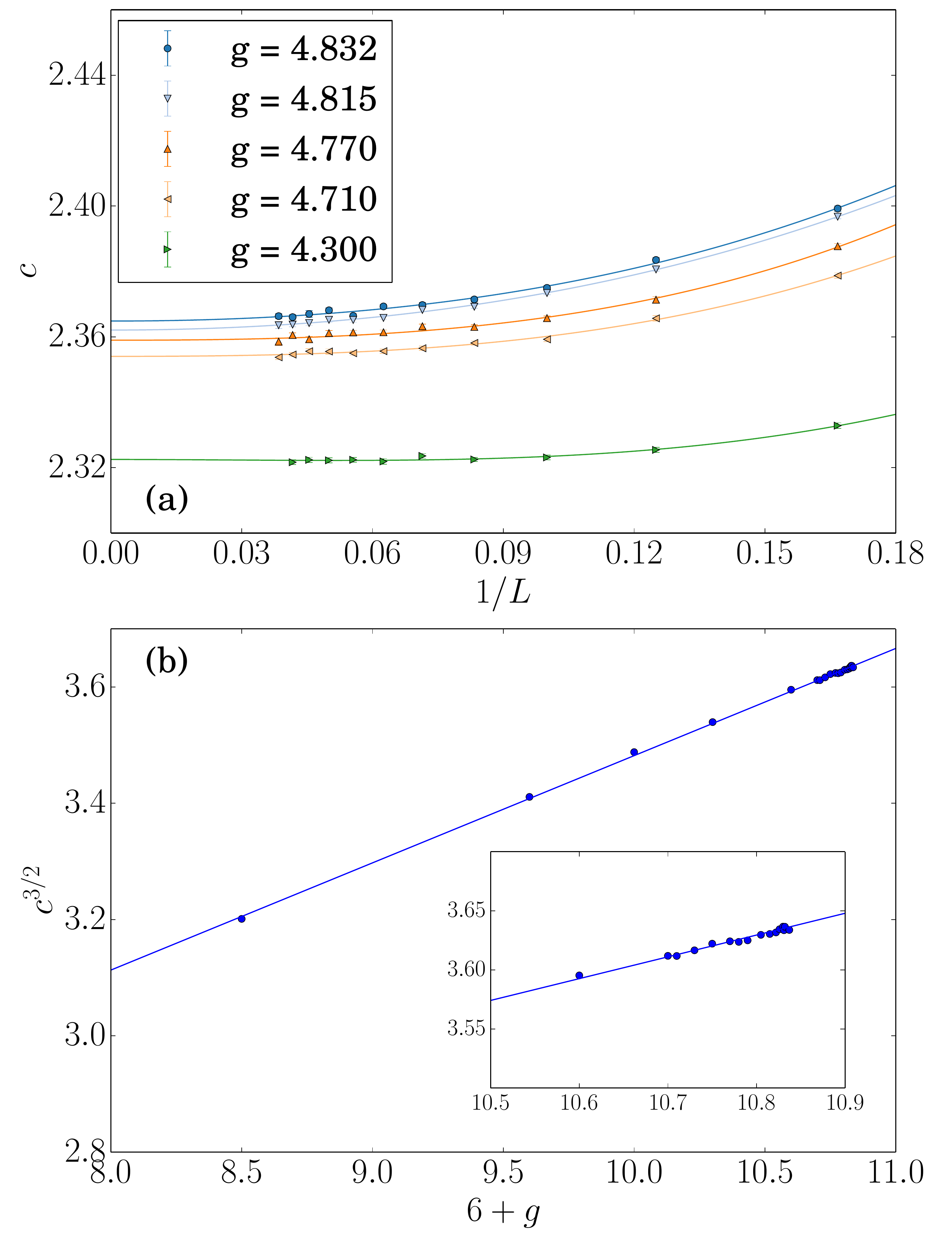}
\caption{(Color online) Calculation of the spin-wave velocity. (a) Velocities 
$c(L)$ obtained for systems of sizes from $L = 6$ to 26 at different values 
of the coupling $g$ and extrapolated using Eq.~(\ref{cexform}). 
(b) Extrapolated velocities $c(g)$ as a function of the microscopic energy 
scale $6 + g$. Error bars are mostly hidden inside the symbols. The solid 
line is a linear fit and the inset magnifies the region close to $g_c$.}
\label{fig:c}
\end{figure}

In Fig.~\ref{fig:c}(a) we show the results for $c(g,L)$ of calculations on 
finite systems of even sizes up to $L = 26$, which we extrapolate to the 
thermodynamic limit using the relation 
\begin{equation}
c(g,L) = c(g) + a(g)/L^2 + b(g)/L^3.
\label{cexform}
\end{equation}
This form is found empirically \cite{Sen2015} to provide a very good 
reproduction of the data and numerical errors due to finite-size effects 
in the critical regime are clearly small. Figure~\ref{fig:c}(b) shows the 
results for the extrapolated spin-wave velocities $c(g)$ of the infinite 
system, by comparing the macroscopic scale $c^{3/2}$ with the microscopic 
quantity $6 + g$ discussed above. The almost perfect linearity demonstrates 
that the two effective energy scales are very closely related, which can 
be expected from the fact that the velocity of a spin excitation depends 
directly on the net interaction of a single spin, and both are perfectly 
valid choices for the normalization of $T_N$. A graph completely analogous 
to Fig.~\ref{fig:tng}, showing a logarithmic correction with the same 
exponent $\hat\tau = 3/11$, is obtained if $c^{3/2}$ is used to normalize 
$T_N(g)$. 

\subsection{Relation between $T_N$ and $m_s$}\label{subsec:6}

In experiment it is often difficult to relate an external control parameter
to the microscopic coupling constants of a model Hamiltonian. In a quantum
antiferromagnet, some aspects of this problem can be circumvented by 
studying the relationship between $T_N$ and $m_s(T = 0)$ directly, without 
reference to the control parameter, $g$. A universal relationship between 
these macroscopic and measurable quantites would be of considerable 
experimental utility in characterizing the nature of critical phenomena 
without recourse to detailed microscopic knowledge of the system parameters 
(such as the pressure dependence of the exchange couplings in TlCuCl$_3$).
Although an experimental test \cite{Merchant2014} of the linear relationship 
between $T_N$ and $m_s$ \cite{Jin2012} indicated satisfactory agreement close 
to the QCP, the issue of how best to normalize $T_N$ was not addressed. We 
use our systematic data spanning the entire QC regime to test the limits of 
linear proportionality and discuss the normalization of $T_N$.

In Ref.~\cite{Jin2012}, where a universal linear relation was found in 
three different models, the authors articulate a mean-field argument based 
on semiclassical considerations for a direct proportionality of $T_N$ to the 
effective spin order gauged by $m_s$ at $T = 0$. In Ref.~\cite{Oitmaa2012}, 
these arguments were elucidated in a field-theory context, where it was 
stated that logarithmic corrections should be negligible for linear 
proportionality to emerge. In fact these arguments can be reduced to 
the statement that it should be possible to treat quantum and thermal 
fluctuations independently, with no mutual interference of their effects 
\cite{Jin2012}. If one considers that mean-field exponents are valid in 
high-dimensional systems $(D > D_c)$ because thermal fluctuations become 
independent of quantum fluctuations when the phase space is sufficiently 
large, then it appears that weak logarithmic corrections could enter the 
relationship of $T_N$ to $m_s$ at $D = D_c$. This possibility, also motivated 
by the (then) unknown form of the logarithmic corrections to $T_N$, was 
investigated directly by QMC simulations for the cubic lattice \cite{Kao2013}, 
but the results were not conclusive (claims concerning the observation of 
logarithmic corrections are not justified by the available data range). Here 
we have presented scaling arguments (Sec.~VIA) and numerical data (Sec.~VIB)
demonstrating that the logarithmic corrections to $m_s$ and $T_N$ have 
precisely the same form, setting their linear relationship in this class of 
system on a far firmer foundation. 

\begin{figure}[t]
\includegraphics[width=\columnwidth]{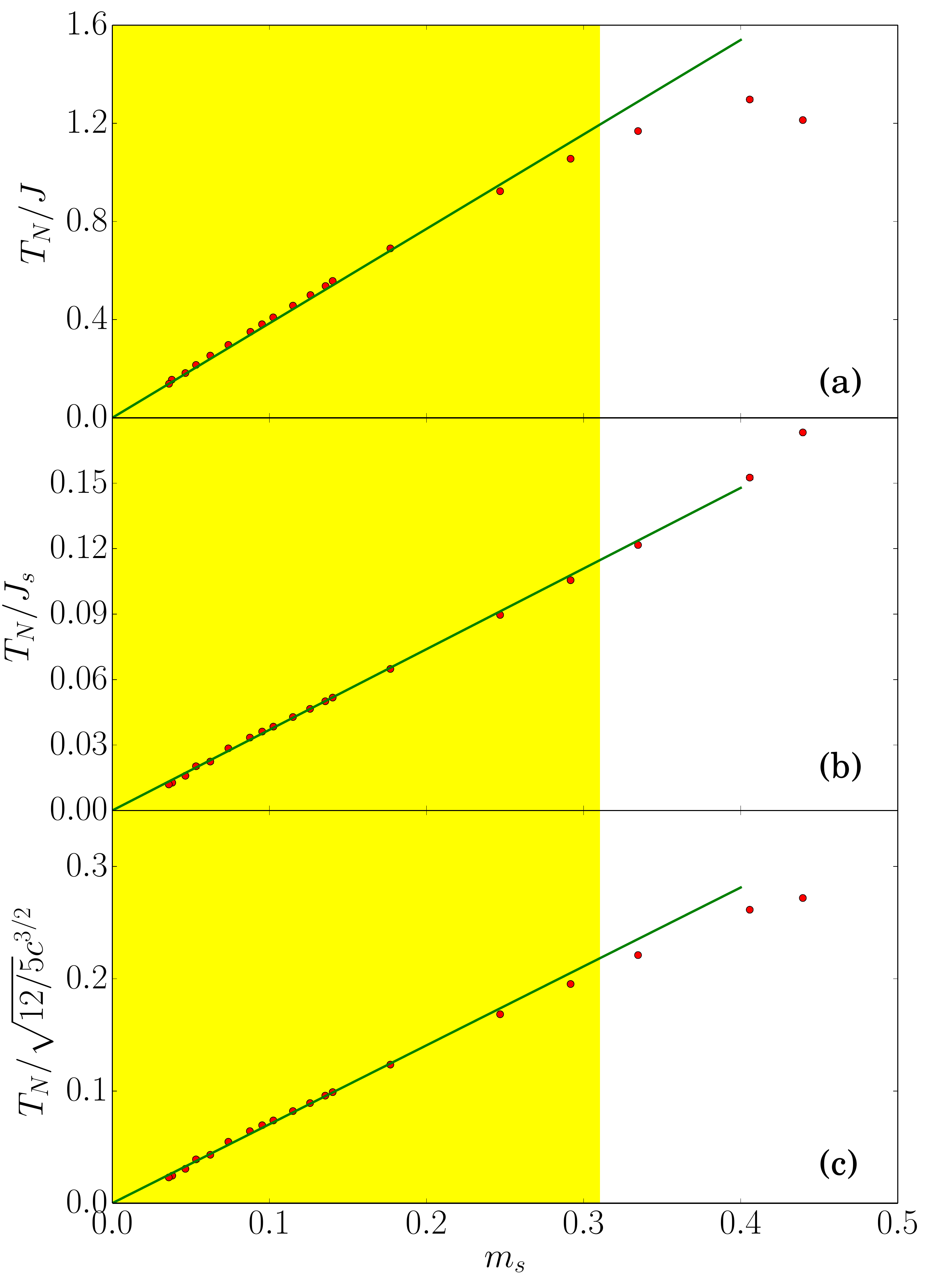}
\caption{(Color online) Relationship between $T_N$ and $m_s$, using different
normalizations of $T_N$; in (a) by the intra-cube coupling $J = 1$, in (b) by 
the sum of couplings $J_s = 6 + g$, and in (c) by the spin-wave velocity in 
the form $\sqrt{12/5} c^{3/2}$. The lines are linear (proportionality) fits
to the small-$m_s$ points and the yellow shaded areas denote the QC regime
determined from $m_s(g)$ in Fig.~\ref{fig:ms2}.}
\label{fig:mstn}
\end{figure}

Our data from Secs.~\ref{sec:ms} and \ref{sec:tn} can be used to probe the 
$T_N(m_s)$ relation in detail and confirm that linearity extends much closer 
to the QCP than previous studies could show. Because our results are for a 
single type of dimerized model, we are not able to address the question of 
a universal prefactor \cite{Jin2012}. However, we are able to make a definite 
statement regarding logarithmic corrections in the relationship between $T_N$ 
and $m_s$. Figure~\ref{fig:mstn} shows our data, taken from Figs.~\ref{fig:msg}
and \ref{fig:tng}, in the form $T_N(m_s)$, with the implicit control parameter 
$g$ effectively eliminated. In panel (a), $T_N$ is simply normalized by the 
energy scale, the interdimer coupling $J = 1$; in panel (b), we have normalized 
$T_N$ by the composite scale $J_s = 6 + g$ (where $g$ can be considered a 
function of $m_s$), as in Fig.~\ref{fig:tng}; in panel (c), we have normalized 
$T_N$ using the correctly scaled spin-wave velocity, $c^{3/2}$. The shaded 
regions again signify our definition of the critical region, based on the 
strict critical scaling of $m_s(g)$ in Fig.~\ref{fig:ms2}. 

As shown in Ref.~\cite{Jin2012}, we find in Fig.~\ref{fig:mstn}(b) that the 
linearity of $m_s$ and $T_N/J_s$ extends well beyond the QC regime; although 
our data were not selected to focus on this region, we do find complete 
agreement with previous calculations where the data overlap. Here we 
demonstrate that the essentially perfect linearity also extends much 
closer to $m_s = 0 = T_N$ ($g \to g_c$), and in particular that it 
remains valid throughout a regime with explicit logarithmic corrections 
in the individual quantities. Although we cannot show that the linear 
relationship extends all the way to the QCP, our data certainly suggest 
that this is the case, i.e.~that it can also be considered a universal 
property of the QC regime. To the extent that linearity of $m_s$ 
and $T_N$ is a consequence of the decoupling of thermal and quantum 
fluctuations, this independence appears to extend from strongly ordered 
systems, where no logarithmic corrections are expected, to the most strongly 
fluctuating QC systems. The linear relationship we have demonstrated verifies 
in full our scaling prediction that the logarithmic corrections to $T_N(g)$
have the same exponent, $\hat\tau = \hat\beta = 3/11$, as $m_s(g)$. We also 
note that the normalization of $T_N$ has no effect (beyond the prefactor) on 
the linear relationship in the QC regime, but that deviations from linearity 
clearly differ at higher $m_s$. 

We close by considering in more detail the case where $T_N$ is normalized by 
$c^{3/2}$ [Fig.~\ref{fig:mstn}(c)], with a view to making quantitative 
comparisons with field-theory predictions \cite{Oitmaa2012}. The explicit 
relationship is 
\begin{equation}
T_N = \gamma {c^{3/2}}\sqrt\frac{{12}}{5} m_s,
\end{equation}
where $\gamma = \langle \phi \rangle/m_s$ is the dimensionful prefactor 
relating the expectation value of the un-normalized field $\phi$ in the 
action to the order parameter $m_s$ of the lattice model. In the 
dimensionless units of our work ($J = 1, \hbar = 1$, \dots), we obtain 
$\gamma = 0.6998 \pm 0.0016$, thereby providing a bridge between the 
quantum field theory and the microscopic lattice Hamiltonian. 
We suggest that this calculation should be repeated for other dimerized 
geometries to test the universality of $\gamma$. It is worth repeating in 
this context the advantages of the double cubic lattice in making the 
spin-wave velocity equal in all three primary axial directions. The 
winding-number method used to extract $c$ in Sec.~\ref{subsec:5} can be 
generalized to anisotropic systems \cite{jiang11b}, but incurs the significant 
complication of altering the aspect ratio of the spatial lattice. Different 
techniques for computing the velocities, such as those based on the 
hydrodynamic relationship among $c$, the spin stiffness, and the magnetic 
susceptibility \cite{Sen2015}, may then be more convenient in practice.

\subsection{Width of the Classical Critical Region}

A key question raised by the experiments on TlCuCl$_3$ \cite{Merchant2014} 
concerns the width of the region close to $T_N$ where classical critical 
scaling applies. It was found that this width, $W \simeq 0.2 T_N$, is 
essentially constant when normalized by $T_N$. We employ scaling arguments 
to show that the normalized width, ${\tilde W} = W/T_N$, should indeed be a 
constant with only a weak logarithmic correction in 3+1 dimensions. 

For fixed temperature $T$, we consider the correlation length, defined in 
terms of the approach of the spin-spin correlation function to its asymptotic 
long-range value, $m_s^2$, when approaching the critical coupling ratio $g(T)$ 
from the ordered side. This quantity has an initial divergence governed by 
the 4D QCP,
\begin{equation}
\xi(T) = \xi_4(T) \sim [g_c(T = 0) - g(T)]^{-\nu_4}, 
\end{equation}
with mean-field exponent $\nu_4 = 1/2$, because the temporal thickness 
$L_\tau$ far exceeds $\xi$ and the system cannot sense its finite temporal 
extent, behaving as at $T = 0$. At the point where $\xi$ reaches $L_\tau$, the 
behavior crosses over to a 3D scaling form, 
\begin{equation}
\xi(T) = \xi_3(T) \sim [g_c(T) - g(T)]^{-\nu_3},
\end{equation} 
where $\nu_3 \approx 0.70$ is the 3D O($3$) exponent. Without logarithmic 
corrections, the temporal length is simply $L_\tau \propto 1/T$ (more 
precisely, $L_\tau = L/c$), but at the upper critical dimension this 
relationship is modified by a logarithmic factor, 
\begin{equation}
L_\tau \sim |\ln(T)|^{\hat q}/T, 
\end{equation} 
which is obtained by generalizing the classical result of Kenna \cite{rkbc}. 
For the 4D O($3$) universality class, $\hat q = 1/4$ \cite{rkbc,rkb}, and 
the correlation length itself also has a logarithmic correction,
\begin{equation}
\xi_4 \sim (g_c - g)^{-\nu_4} |\ln (g_c - g)|^{\hat\nu}, 
\end{equation} 
with $\hat\nu = 5/22$. The quantum-classical crossover taking place when 
$\xi \approx L_\tau$ therefore corresponds to 
\begin{equation}
|\ln(T)|^{\hat q}/T \sim [g_c - g)]^{-\nu_4} |\ln (g_c - g)|^{\hat\nu}, 
\end{equation}
which, by converting to a temperature-dependence and keeping only the leading 
logarithm, yields the crossover temperature 
\begin{equation}
T^*(g) \sim (g_c - g)^{1/2} |\ln (g_c - g)|^{\hat q - \hat\nu}.
\end{equation}
Using our result for $T_N(g)$ [Eq.~(\ref{tnlog322})], the width of the 
classical critical region on the ordered side of the transition is therefore 
\begin{equation}
{\tilde W}(g) = \frac{T_N(g) - T^*(g)}{T_N(g)} \sim 1 - a|\ln (g_c - g)|^{(\hat 
q - \hat\nu)/\hat \tau},
\end{equation}
with a constant $a$, whose calculation requires further considerations, and 
a small exponent 
\begin{equation}
\frac{\hat q - \hat\nu}{\hat \tau} = \frac{1/4 - 5/22}{3/11} = 1/12, 
\end{equation}
on the logarithm. This very weak dependence explains the near-constant behavior 
found for TlCuCl$_3$ \cite{Merchant2014}. On general grounds we expect the 
width of the classical critical regime on the other side of the transition 
to scale in the same way. 

\section{Summary}
\label{sec:summary}

We have provided a direct and non-perturbative verification of the existence 
and nature of multiplicative logarithmic corrections to scaling at the 
quantum phase transition for three-dimensional dimerized quantum Heisenberg 
antiferromagnets. These systems correspond to the $\phi^4$ field theory of 
an O($3$) quantum field in $3 + 1$ dimensions, which is the upper critical 
dimension ($D_c = 4$) for all models with O($N$) universality. With the 
exception of the Ising model ($N = 1$) \cite{Kenna1994}, no such 
demonstration exists to date, despite a significant body of analytical and 
numerical work on quantum criticality in dimerized quantum antiferromagnets. 

Our results are obtained from large-scale quantum Monte Carlo 
calculations based on state-of-the-art simulation techniques and 
detailed finite-size-scaling analysis. These enabled us to extract the 
precise logarithmic corrections to the leading critical properties at 
the quantum phase transition from a non-magnetic state of dominant dimer 
correlations to a N\'eel-ordered antiferromagnetic state. Specifically, we 
have obtained the multiplicative logarithmic corrections to the mean-field  
behavior of the order parameter, the zero-temperature staggered magnetization 
($m_{s}$), on the control parameter, the coupling ratio $g$. We have 
verified that these are governed by an exponent ${\hat \beta} = 3/11$, 
a value we specify with numerical (statistical) precision under $3\%$, 
matching precisely the prediction of perturbative renormalization-group 
calculations \cite{Zinn-Justin2002,rkbc}. 

No prediction was previously available for the analogous logarithmic 
correction to the N\'eel temperature, $T_N$. We have implemented a 
scaling Ansatz exploiting the known logarithmic corrections of other 
physical quantities to obtain its form. Our prediction is that $T_N$  
has exactly the same exponent in its logarithmic correction, ${\hat \tau}
 = 3/11$, as the order parameter, and our numerical results for $T_N(g)$ 
are in excellent agreement. We have thereby established an exact linearity 
between $T_{N}$ and $m_{s}$ throughout the quantum critical regime. We have 
also demonstrated a different kind of logarithmic correction, in the 
size-dependence of the staggered magnetic susceptibility at the 
four-dimensional quantum critical point, where we verify the predicted 
$N$-independent scaling form \cite{Kenna2004}. 

The numerical task of finding logarithmic corrections is not a 
straightforward one. We have established that the appropriate scaling 
regime is $|g - g_c|/g_c \lesssim 0.2$. Within this region, obtaining 
reliable evidence for logarithmic corrections is critically dependent on 
having many high-precision data points at very small values of $|g - g_c|$, 
which mandates accurate calculations at large system sizes. After 
establishing the location of the critical point to approximately one part 
in $10^5$, $g_c = 4.83704(6)$, we were able to obtain highly accurate 
extrapolations of the physical observables for coupling ratios as close 
to $g_c$ as 4.834, i.e.~with $|g - g_c|/g_c \simeq 0.0006$. This required 
working with linear system sizes as large as $L = 48$, meaning a system 
containing $N = 2L^{3} = 221184$ interacting spins, and at temperatures as 
low as $T = 1/(2L) = 1/96$. From this perspective, it becomes obvious why 
previous studies \cite{Jin2012,tbk07,Kao2013}, with only a handful of data 
points in the quantum critical regime (none closer than $|g - g_c|/g_c = 
0.02$) were not able to find any meaningful evidence for logarithmic 
corrections. 

Our results are directly relevant to the pressure-induced quantum phase 
transition in TlCuCl$_3$ \cite{rrs2004,Ruegg2008,Merchant2014}. Detailed 
experiments on this material by elastic and inelastic neutron scattering 
have measured the staggered magnetization, the N\'eel temperature, the 
gap of the quantum disordered phase, and the magnetic excitation spectrum 
on both sides of the transition. On the assumption that the leading 
dependence of the control parameter (the ratio of antiferromagnetic 
superexchange parameters) is linear in the applied pressure, both $m_s$ 
and $T_N$ show good mean-field exponents and a close linear relation over 
much of the accessible pressure range \cite{Merchant2014}. On the grounds 
that the available data follow mean-field scaling around the quantum critical 
point, it cannot be argued that they provide any evidence for logarithmic 
corrections, although the size of the experimental errors and the shortage 
of data very close to the QCP certainly mean they cannot be excluded.

It has been argued very recently \cite{Scammell2015}, based on a 
field-theoretic treatment, that the apparent suppression of $m_s$ and 
$T_N$ visible in the experimental data for TlCuCl$_3$ rather far from 
the QCP (at pressures 2-4 times the critical pressure) arises due to 
logarithmic scaling of the coupling constant. In Ref.~\cite{Merchant2014}
it was assumed that these effects are in fact a consequence of departures 
from the quantum critical scaling regime, evident also in the violation 
of linearity between $m_s$ and $T_N$ beyond the point where the order 
parameter is 60\% of the classical moment. Although a direct comparison 
with our results is not possible without a microscopic treatment of the 
relationship between the applied pressure and the control parameter, a 
similar downturn is visible, and better described by including the 
multiplicative logarithmic corrections, beyond $|g - g_c|/g_c \approx 0.05$ 
in our Figs.~\ref{fig:ms} and \ref{fig:tng}. It is also tempting to 
relate the $T_N (m_s)$ curve of TlCuCl$_3$ \cite{Merchant2014} to our 
Fig.~\ref{fig:mstn}(b), where the extended linear regime is followed by 
an upturn deep inside the N\'eel phase, which was interpreted \cite{Jin2012} 
as the breakdown of the quantum-thermal decoupling (see below) due to a large 
density of thermally excited magnons when $T_N$ is high. Finally, we have 
also provided a theoretical explanation for the shape of the classical 
critical scaling ``fan'' around $T_N(p)$ observed in TlCuCl$_3$ 
\cite{Merchant2014}, by showing that its width scales linearily with 
$T_N$, modified by a logarithmic correction with a very small exponent of 
$1/12$, which would vary extremely weakly over the experimental pressure 
window.

Although many dimerized $S = 1/2$ systems with antiferromagnetic interactions 
are known, and many field-induced quantum phase transitions have been studied, 
few have yet been found to be close to quantum critical points at zero field 
under pressure. Our results shed light on the experimental challenges inherent 
in finding logarithmic corrections, but also provide evidence that their
detection is actually possible. While important theoretical questions 
remain to be addressed in lower dimensions, logarithmic corrections are of 
little relevance away from $D_c$. Another challenge for both experiment and 
numerical simulation would be to investigate the exponents and corrections 
for different $N$, meaning for systems of Ising and XY spins. A related 
experimental possibility would be to realize the $N = 2$ situation in a 
gas of ultra-cold bosons on an optical lattice. The unfrustrated dimerized 
antiferromagnet is a bipartite lattice and thus can be treated exactly as a 
system of hard-core dimer bosons, with the dimerized phase corresponding to 
the Mott insulator and the antiferromagnet to the superfluid (a state of 
long-range inter-site coherence); the symmetry broken is U(1), which is 
equivalent to XY. Although these experiments have not yet been realized in 
sufficiently large three-dimensional gases of cold bosons, the very fine 
parameter control possible in cold-atom systems offers another candidate 
route for the experimental observation of logarithmic corrections to scaling. 

One of the points made in Ref.~\cite{Merchant2014} was that, although 
quantum critical phenomena are universal, obeying scaling forms determined 
only by macroscopic properties of the system such as the dimensionality and 
the symmetry of the order parameter, their experimental observation depends 
crucially on non-universal prefactors. For quantum critical excitations, this 
is the ratio of the width of an excitation to its energy, and is a quantity 
determined entirely by microscopic details. For both static and dynamic 
properties, the key figure of merit is the width of the quantum critical 
regime, and for this we have obtained a quantitative result not previously 
available by any other technique, $|g - g_c|/g_c \lesssim 0.2$. In as much 
as one may generalize from the dimensionality and geometry of the double 
cubic lattice, this 20\% criterion dictates the necessary proximity to the 
quantum critical point for the observation of strict quantum critical 
scaling, including logarithmic corrections. 

Our demonstration of linearity between $m_s$ and $T_N$ in the $(3 + 1)$D 
Heisenberg antiferromagnet lies beyond any results previously predicted by 
analytical methods. What we have demonstrated explicitly for several 
quantities is the presence of expected logarithmic corrections, but their 
cancellation between $m_s$ and $T_N$ was not anticipated. However, in 
parallel to our scaling argument for the logarithmic corrections to $T_N$, 
Scammell and Sushkov have recently arrived at the same conclusion from a 
different starting point \cite{Scammell2015}. A key outstanding question 
is whether the linearity of $T_N(m_s)$ is in fact a more fundamental property 
of the system than arguments made at the semiclassical and mean-field levels 
suggest. Qualitatively, the origin of linearity is thought \cite{Jin2012} to 
lie in the effective decoupling of the classical and quantum fluctuations, 
which is applicable for all coupling ratios both outside and inside the QC 
regime. Its observation here implies the enduring independence of quantum 
and thermal fluctuations at the O($N$) transition with $D = D_c$ for any 
$N$. Efforts to study the relationship between the $T = 0$ order parameter 
and the critical temperature in systems with different universality classes 
would shed light on this matter.

\begin{acknowledgments}
We thank A. Honecker, R. Kenna, D. Lin, H. Scammell, and O. P. Sushkov 
for helpful discussions. This work was supported by the National 
Thousand-Young-Talents Program of China (YQQ and ZYM), by the National 
Natural Science Foundation of China under Grant No.~11174365, and by the 
National Basic Research Program of China under Grant No.~2012CB921704 (BN). 
The simulations were carried out at the National Supercomputer Center in 
Tianjin on the platform TianHe-1A. AWS was supported by the NSF under 
Grant No.~DMR-1410126, the Simons Foundation, and by the Center for 
International Collaboration at the Institute of Physics of the Chinese 
Academy of Sciences.
\end{acknowledgments}

\appendix*

\section{Crossing-point scaling in the presence of logarithmic corrections}
\label{app:crossings}

According to the general hypothesis of finite-size scaling (FSS), 
verified by renomalization-group techniques (for a review see 
Ref.~\cite{Pelissetto02}), the dependence on system size of 
a physical quantity in the neighborhood of a critical point can be 
described by the function
\begin{equation}
Q(t,L) = L^{\kappa/\nu} [f(\xi/L) + O(L^{-\omega},\xi^{-\omega})],
\label{eq:app_FSS}
\end{equation}
where $t$ is the distance to the critical point, i.e.~$t = |T - T_c|$ for a 
classical or $t = |g - g_c|$ for a quantum phase transition. $\kappa$ is the 
critical exponent for the quantity in question in the thermodynamic limit, 
$Q(t) \sim |t|^{-\kappa}$, and the subleading exponent $\omega$ originates 
from an irrelevant scaling field. In a fully rigorous treatment, $L$ in 
$f(\xi/L)$ should be replaced by $\xi_L(0)$, which is the finite-size 
correlation length at the critical point ($t = 0$) and thus the relevant 
length scale for FSS. In the absence of logarithmic corrections, $\xi_L(0) 
\sim L$.

The leading term in Eq.~(\ref{eq:app_FSS}) is the asymptotic FSS and the 
second term expresses the correction to scaling. For the Binder ratio, $Q(g,L)
 = R_{2}(g,L)$, which is a dimensionless ``invariant,'' the asymptotic scaling 
has exponent $\kappa = 0$. Correction terms remain present, and at the 
crossing point $t^{*}$ for two system sizes $L_1$ and $L_2$ one has 
\begin{equation}
R_2(t^{*},L_1) = R_2(t^{*},L_2).
\label{eq:app_R2}
\end{equation}
Without logarithmic corrections, 
\begin{equation}
f(\xi/L) = h(t L^{1/\nu})
\label{eq:app_asymptotic}
\end{equation}
and hence 
\begin{equation}
R_2(t,L) = a + b t L^{1/\nu} + cL^{-\omega} + \dots
\label{eq:app_R2_expan}
\end{equation}
The crossing point, $t^{*}$, can be determined for ($L_1$, $L_2$) as
\begin{equation}
t^{*} \sim \frac{1 - s^{-\omega}}{s^{1/\nu} - 1} L_1^{-\omega-1/\nu},
\label{eq:app_tcross}
\end{equation}
where $s = L_2/L_1$, and is either constant ($L_2 = a L_1$ with $a > 0$) or 
approaches unity ($L_2 = L_1 + 2$) as the system size goes to infinity. 

If the logarithmic correction to the correlation length is taken into 
consideration,
\begin{equation}
\xi \sim t^{-\nu} |\ln t|^{\hat \nu}
\end{equation}
and, according to Refs.~\cite{rkbc,rkb},
\begin{equation}
\xi_L(0) \sim L \ln^{\hat q} L 
\label{eq:app_fsslog}
\end{equation}
is now the relevant FSS length scale. On substituting Eq.~(\ref{eq:app_fsslog})
into both the asymptotic FSS term $f(\xi/L)$ and the subleading term 
$L^{-\omega}$, Eq.~(\ref{eq:app_tcross}) becomes
\begin{equation}
t^{*} \ln^{{\hat \nu}/\nu} t^{*} \sim \frac{1 - s'^{-\omega}}{s'^{1/\nu} - 1} 
\xi_{L_1}^{-\omega-1/\nu}(0)
\label{eq:app_tcross_log}
\end{equation}
where
\begin{equation}
s' = \frac{\xi_{L_2}(0)}{\xi_{L_1}(0)} = \frac{L_2 \ln^{\hat q}(L_2)}{L_1 
\ln^{\hat q}(L_1)},
\end{equation}
which also approaches a constant as $L_1,L_2 \to \infty$.

There is no straigtforward inversion of Eq.~(\ref{eq:app_tcross_log}) to 
obtain an exact expression for $t^{*}$. However, for the leading logarithmic 
correction it is sufficient to substitute Eq.~(\ref{eq:app_tcross}) into the 
logarithmic part of (\ref{eq:app_tcross_log}), which yields
\begin{equation}
\ln t^{*} \sim c + \ln \left (\frac{1 - s^{-\omega}}{s^{1/\nu} - 1} \right ) - 
(\omega + 1/\nu) \ln L,
\end{equation}
a quantity approximately proportional to $\ln L $ when $L$ is large (consider 
$s = {(L+2)}/{L} = 1 + {2}/{L}$). The leading scaling behavior, obtained on 
replacing $\ln t^{*}$ in Eq.~(\ref{eq:app_tcross_log}) by $\ln L $, is
\begin{eqnarray}
t^{*} & \sim & \frac{1 - s'^{-\omega}}{s'^{1/\nu} - 1} L^{-\omega-1/\nu} 
\ln^{-\hat q \omega - \hat q/\nu} L \ln^{\hat\nu/\nu} L \nonumber \\
& \sim & \frac{1 - s'^{-\omega}}{s'^{1/\nu} - 1} L^{-\omega-1/\nu} \ln^{\hat c} L, 
\label{eq:app_t}
\end{eqnarray}
where the exponent $\hat c$ is given by
\begin{equation}
\hat c = \frac{\hat\nu - \hat q}{\nu} - \hat q \omega.
\label{eq:app_lambda1}
\end{equation}
Replacing $t$ by $g - g_c$ and taking the large-$L$ limit such that $s'\to 1$, 
Eq.~(\ref{eq:app_t}) yields 
\begin{equation}
g_{c}(L) = g_{c} + aL^{-(1/\nu+\omega)} \ln^{\hat c} L,
\end{equation}
which is Eq.~(\ref{eq:gcpowerformlog}) in Sec.~\ref{sec:qcpscaling}.
However, if there is no logarithmic correction to the subleading term 
$L^{-\omega}$, the second term in Eq.~(\ref{eq:app_lambda1}) is absent, and 
\begin{equation}
\hat c = \hat \lambda = \frac{\hat\nu - \hat q}{\nu}.
\end{equation}

\end{document}